\documentclass[acmsmall]{acmart} %

\AtBeginDocument{%
  \providecommand\BibTeX{{%
    \normalfont B\kern-0.5em{\scshape i\kern-0.25em b}\kern-0.8em\TeX}}}

\setcopyright{acmlicensed}
\copyrightyear{2024}
\acmYear{2024}
\acmDOI{XXXXXXX.XXXXXXX}

\acmJournal{TOSEM}

\usepackage{amsmath,amsfonts}
\usepackage{graphicx}
\usepackage{textcomp}

\usepackage{caption}
\usepackage{blindtext}
\usepackage{multicol}
\usepackage{xcolor}
\usepackage{tikz}
\usepackage{listings}
\usepackage{enumitem}
\usepackage{hyperref}
\usepackage{amsfonts}
\usepackage{wrapfig}
\usepackage{tcolorbox}
\usepackage{lipsum}
\usepackage{xspace} %
\usepackage{subcaption} 
\usepackage{adjustbox}
\usepackage{fancybox}

\usepackage[noend]{algpseudocode}
\usepackage[ampersand]{easylist}
\newtheorem{definition}{Definition}

\usepackage{colortbl}
\usepackage[normalem]{ulem}  %
\useunder{\uline}{\ul}{}  %
\usepackage{booktabs} %
\usepackage{multirow} %

\definecolor{main}{HTML}{5989cf}    %
\definecolor{sub}{HTML}{cde4ff}     %

\newtcolorbox{boxB}{
    fontupper = \bf\color{main}\footnotesize, %
    boxrule = 0.5pt,
    colframe = main,
    rounded corners,
    arc = 5pt   %
}

\newtcolorbox{boxD}{
    fontupper = \small, 
    colback = sub, 
    colframe = main, 
    boxrule = 0pt, 
    toprule = 2pt, %
    bottomrule = 2pt %
}

\newtcolorbox{boxH}{
    fontupper = \small, 
    colback = sub, 
    colframe = main, 
    boxrule = 0pt, 
    leftrule = 6pt %
}

\newtcolorbox{boxG}{
    enhanced,
    boxrule = 0pt,
    colback = sub,
    borderline west = {1pt}{0pt}{main}, 
    borderline west = {0.75pt}{2pt}{main}, 
    borderline east = {1pt}{0pt}{main}, 
    borderline east = {0.75pt}{2pt}{main}
}    

\newtcolorbox{boxK}{
    fontupper = \small,
    sharpish corners, %
    boxrule = 0pt,
    toprule = 1.0pt, %
}

\definecolor{MidnightBlue}{HTML}{006895}

\newcommand*\circled[1]{\tikz[baseline=(char.base)]{
            \node[shape=circle,draw,inner sep=0.5pt] (char) {#1};}}

{
}

\newcommand{\ie}{\textit{i.e.,}\xspace}
\newcommand{\eg}{\textit{e.g.,}\xspace}

\newcommand{\etal}{et al.\xspace}

\newcommand{\aka}{\textit{a.k.a.}\xspace}   

\newcommand{\approach}{AST\textit{rust}\xspace}

\newcommand{\llms}{LLMs\xspace}
\newcommand{\llm}{LLM\xspace}

\newcommand{\asc}{\textit{SC}\xspace}
\newcommand{\ascs}{\textit{SCs}\xspace}
\newcommand{\ntp}{\textit{TLP}\xspace}
\newcommand{\scm}{\textit{SCM}\xspace}

\newcommand{\galeras}{\textit{SyxTestbed}\xspace}%

\newcommand{\gptI}{\textit{gpt-3 [125M]}\xspace}
\newcommand{\gptII}{\textit{gpt-3 [1.3B]}\xspace}
\newcommand{\gptIII}{\textit{gpt-3 [2.7B]}\xspace}

\newcommand{\codegenII}{\textit{codegen-nl [2B]}\xspace}

\newcommand{\multiI}{\textit{multi-lang [110M]}\xspace}

\newcommand{\multiIII}{\textit{multi-lang [2B]}\xspace}

\newcommand{\monoI}{\textit{mono-lang [110M]}\xspace}
\newcommand{\monoII}{\textit{mono-lang [1.5B]}\xspace}

\newcommand{\monoIIII}{\textit{mono-lang [2B]}\xspace}

\newcommand{\nlgpt}{\textit{NL GPT-3}\xspace}
\newcommand{\nlcodegen}{\textit{NL Codegen}\xspace}
\newcommand{\monolang}{\textit{Mono-Language-Type}\xspace}
\newcommand{\multilang}{\textit{Multi-Language-Type}\xspace}

\newcommand{\secref}[1]{Sec.~\ref{#1}\xspace}
\newcommand{\figref}[1]{Fig.~\ref{#1}\xspace}
\newcommand{\tabref}[1]{Tab.~\ref{#1}\xspace}

\newcommand{\surveyControl}{\textit{$U_{CTR}$}\xspace}

\newcommand{\surveySequence}{\textit{$U_{SEQ}$}\xspace}
\newcommand{\surveyAstBased}{\textit{$U_{AST}$}\xspace}

\newcommand{\surveyAstPartial}{\textit{$U_{AST[p]}$}\xspace}
\newcommand{\surveyAstComplete}{\textit{$U_{AST[c]}$}\xspace}

\begin{document}

\title{Towards More Trustworthy and Interpretable LLMs\\ for Code through Syntax-Grounded Explanations}
\author{David N. Palacio}
\email{danaderpalacio@wm.edu}
\orcid{0001-6166-7595}
\affiliation{%
  \institution{William \& Mary}
  \city{Williamsburg}
  \state{Virginia}
  \country{USA}
  \postcode{23185}
}

\author{Daniel Rodriguez-Cardenas}
\orcid{0002-3238-1229}
\affiliation{%
  \institution{William \& Mary}
  \city{Williamsburg}
  \state{Virginia}
  \country{USA}
  \postcode{23185}
  }
\email{dhrodriguezcar@wm.edu}

\author{Alejandro Velasco}
\affiliation{%
  \institution{William \& Mary}
  \city{Williamsburg}
  \state{Virginia}
  \country{USA}
  \postcode{23185}
}
\email{svelascodimate@wm.edu}
\orcid{0002-4829-1017}

\author{Dipin Khati}
\affiliation{%
 \institution{William \& Mary}
 \city{Williamsburg}
 \state{Virginia}
 \country{USA}
 \postcode{23188}
 }
 \email{dkhati@wm.edu}

\author{Kevin Moran}
\affiliation{%
  \institution{University of Central Florida}
  \city{Orlando}
  \state{Florida}
  \country{USA}}
\email{kpmoran@ucf.edu}

\author{Denys Poshyvanyk}
\affiliation{%
  \institution{William \& Mary}
  \city{Williamsburg}
  \state{Virginia}
  \country{USA}
  \postcode{23185}}
\email{dposhyvanyk@wm.edu}

\renewcommand{\shortauthors}{Palacio, et al.}

\begin{abstract}
  Trustworthiness and interpretability are inextricably linked concepts for \llms. The more interpretable an \llm is, the more trustworthy it becomes. 
However, current techniques for interpreting \llms when applied to code-related tasks largely focus on accuracy measurements, measures of how models react to change, or individual task performance instead of the fine-grained explanations needed at prediction time for greater interpretability, and hence trust. 
To improve upon this status quo, this paper introduces \approach, an interpretability method for \llms of code that generates explanations grounded in the relationship between model confidence and syntactic structures of programming languages.
\approach explains generated code in the context of \textit{syntax categories} based on Abstract Syntax Trees and aids practitioners in understanding model predictions at \textit{both} local (individual code snippets) and global (larger datasets of code) levels.
By distributing and assigning model confidence scores to well-known syntactic structures that exist within ASTs, our approach moves beyond prior techniques that perform token-level confidence mapping by offering a view of model confidence that directly aligns with programming language concepts with which developers are familiar. 

To put \approach into practice, we developed an automated visualization that illustrates the aggregated model confidence scores superimposed on sequence, heat-map, and graph-based visuals of syntactic structures from ASTs. 
We examine both the \textit{practical benefit} that \approach can provide through a data science study on 12 popular LLMs on a curated set of GitHub repos and the \textit{usefulness} of \approach through a human study. 
Our findings illustrate that there is a \textit{causal} connection between learning error and an LLM's ability to predict different syntax categories according to \approach~-- illustrating that our approach can be used to interpret model \textit{effectiveness} in the context of its syntactic categories. Finally, users generally found \approach's visualizations useful in understanding the trustworthiness of model predictions.

\end{abstract}



\maketitle

\section{Introduction}\label{sec:introduction}

The proliferation of open-source software projects and rapid scaling of transformer-based Large Language Models (\llms) has catalyzed research leading to the increased effectiveness of automated Software Engineering (SE) tools. \llms have demonstrated considerable proficiency across a diverse array of generative SE tasks~\cite{Chen2021EvaluatingCode, watson2020dl4se}, including, but not limited to, code completion ~\cite{Raychev2014CodeCW,MSR-Completion}, program repair ~\cite{Chen2019sequencer,ahmad_unified_2021}, and test case generation ~\cite{Watson:ICSE20}. Current research in both designing \llms for code and applying them to programming tasks typically makes use of existing \textit{benchmarks} (\eg CodeSearchNet~\cite{husain2019codesearchnet}, or HumanEval~\cite{chen_evaluating_2021}) and \textit{{canonical metrics}} (by \textit{canonical}, we refer to metrics that reflect an \textit{aggregate performance} across many model predictions, for example, percentage accuracy). These canonical metrics have been adapted from the field of Natural Language Processing (NLP) to evaluate the performance of deep code generation models.

Recent work has illustrated the limitations of benchmarks such as HumanEval~\cite{liu2023code} and there has been growing criticism of canonical metrics within the NLP community due to the lack of an \textit{interpretable context} that allows for a deeper understanding of LLMs' predictions or outputs~\cite{molnar2019interpret,Kim2018InterpretabilityTCAV,wan_what_2022,liu_reliability_2023, doshi-velez_towards_2017}. While code-specific metrics such as CodeBLEU~\cite{Ren2020codebleu} may provide more robust aggregate pictures of model accuracy, they cannot provide the fine-grained context required to truly explain model predictions. The general lack of widely adopted interpretability or explainability tools is a barrier to the adoption of any deep learning model, and in particular LLMs of code, as practitioners are skeptical of models' trustworthiness~\cite{lo_trustworthy_2023}. This deficiency largely stems from the fact that such benchmarks and canonical metrics are often aimed at evaluating functional correctness or standard performance of generated code \textit{at a glance}. That is, the evaluation is reduced to a single aggregate metric in which relevant information related to individual predictions is obfuscated~\cite{burnell_rethink_2023}.

\begin{wrapfigure}{l}{0.53\textwidth}
		\centering
         \vspace{-1em} 
		\includegraphics[width=0.55\textwidth]{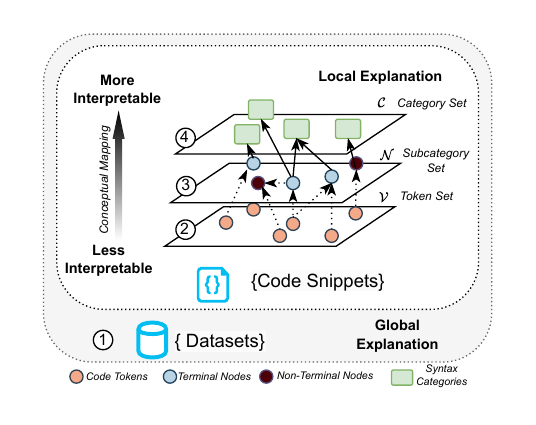}
        \vspace{-3em} 
		\caption{The Conceptual Framework of Syntax-Grounded Interpretability}
        \Description{Sytanx-Grounded Interpretability Overview}
        \label{fig:overview}
\end{wrapfigure}

Methods for \textit{interpreting} and \textit{trusting} \llms for code are inextricably linked. A trustworthy \llm for code requires some degree of interpretability of its predictions, such that model behavior can be understood at a fine-grained enough level to judge which parts of the output are correct or not, and why. The more interpretable an LLM for code is, the higher the confidence and trust in the deployment and use of the model~\cite{Doshi-Velez2018ConsiderationsLearning, ji2024ai}. Notably, interpretability has been identified as an important component for enhancing trustworthiness in various studies ~\cite{lundberg2017unified,weller2019transparency,Liao_2020}. When evaluating trustworthiness, a clear understanding of how and why a model reaches specific predictions is critical. This transparency not only addresses challenges related to uncertainty and the potential for bugs or vulnerabilities but also plays a pivotal role in transforming a model perceived as untrustworthy into one deemed as reliable ~\cite{ribeiro2016why}. %

We assert that a \llm for code is \textit{interpretable}, and hence more trustworthy, if the reasoning behind its predictions is easy for a practitioner to comprehend. In other words, a useful interpretability technique must provide a \textit{conceptual mapping} between descriptions of a model's reasoning process and concepts inherently understood by programmers. In this paper, we explore the possibility of using a model's \textit{confidence} in its predictions as a proxy for describing its reasoning process and develop a technique, which we call \textbf{\approach} that automatically \textit{aligns} and \textit{clusters} model confidence measures with groups of tokens based on syntactic categories derived from Abstract Syntax Trees (ASTs) that we call \textit{Syntax Categories} (\ascs). This method enables a fine-grained understanding of the correctness of model predictions rooted in \textbf{syntax-grounded explanations}. As illustrated by the overview of our approach in Fig.~\ref{fig:overview}, \approach enables two different granularities of interpretability, \textit{local explanations} at the code snippet level, and \textit{global explanations} for large collections of code. %
\approach also makes two main contributions: (i) a statistical technique for aligning and aggregating confidence scores to syntactic code structures of different granularities, and (ii) an automated technique for generating visualizations of these aligned confidence scores. At the \textit{local} level these visualizations take the form of model confidence scores overlaid on both sequence and graph-based illustrations of ASTs and different syntactic structures. At the global level, these take the form of a heat map with confidence values clustered around higher-level syntactic categories. An example of the type of explanation that a developer may derive from \approach's visualizations is as follows, ``\textit{The model's prediction of the type of the} \texttt{\small character} \textit{parameter may be incorrect due to low confidence.}'' %

Grounding explanations of model confidence in code syntax provides an informative context to practitioners allowing for interpretability. This is due to the fact that code semantics and syntax are tightly coupled. That is, descriptions of code meaning, or semantics, are often \textit{grounded} in syntax. For instance, consider the following example of a developer describing program behavior in \texttt{\small numpy} in which the description of functionality is grounded in terms of data structures, \textit{``Convert an \uline{array} representing the coefficients of a Legendre series,''}\footnote{\url{https://github.com/numpy/numpy/blob/main/numpy/polynomial/legendre.py\#L152C5-L152C72}} where the underlined word refers explicitly to the syntactic category of a data structure. One may ask \textit{``why not ground explanations in code semantics directly?''} However, such semantic-based grounding is difficult to achieve, as it requires reasoning among model confidence, input code, predicted code, and widely variable interpretations of code meaning -- leading to the potential for \textit{incorrect} explanations that would undermine a technique ultimately meant to build trust. However, as we illustrate in this paper, it is possible to directly map measures of model confidence to different syntactic categories of code, providing a \textit{statistically sound} method of understanding the potential correctness of model predictions rooted in concepts that developers can easily understand. 

We explore the \textit{practical benefit} of \approach through a large-scale data science study examining the relationship between model effectiveness and global explanations and evaluate the \textit{usefulness} of our method through a human study targeted at local explanations of code snippets using \approach's different visualizations. The context of our empirical evaluation includes 12 popular \llms for code and a curated set of code taken from recent commits of the 200 most popular Python projects on GitHub. Using a carefully crafted causal inference study, our analysis illustrates \textit{causal} connections between learning error and a model's ability to predict different syntax categories according to \approach~-- showing that our approach can be used to interpret model \textit{effectiveness} in the context of its syntactic categories. Our human study included 27 participants who examined code snippets completed by GPT 3 and one of four of \approach's visualization techniques for local explanations. Our results illustrate that developers generally found \approach and its visualizations useful in understanding model predictions.

The results of our studies illustrate that mapping token-level predictions of \llms to segregated Syntax Categories are of considerable practical benefit to SE researchers and practitioners because it allows them to \textbf{interpret} and \textbf{trust} parts of generated code based on the structural functionality, which contextualizes model predictions beyond the canonical evaluation (\ie measuring intrinsic and extrinsic metrics). We hope other researchers build upon our method to create new types of interpretability techniques in the future, and we provide an online appendix with the code for \approach, and our data and experimental infrastructure to facilitate replication~\cite{AnonyRepoASTrust24}.

\section{Background \& Related Work}
\label{sec:background}

In this section, we present background on interpretability and trustworthiness as complementary terms for generating syntax-grounded post hoc (\eg generated after training) explanations for \llms of code. 

\textbf{Interpretability.} The brittleness of LLMs can be formulated as an \textit{incompleteness} in problem formalization ~\cite{Doshi-Velez2017TowardsLearning}, which means that it is insufficient that models only infer predictions for certain tasks (the \textbf{what?}). The models must also explain how they arrive at such predictions (the \textbf{why?}). To mitigate such incompleteness in problem formalization, the field of \textit{interpretability} has risen to encompass techniques and methods that aim to solve the \textit{why} question. Although authors in this field generally use the terms \textit{explainability} and \textit{interpretability} interchangeably, these definitions are inconsistent throughout the literature~\cite{flora_comparing_2022}. We distinguish between the terms to avoid confusion with the purposes of our approach. We will use \textit{explainability} for methods whose goal is to understand how a \llm operates and comes to a decision by exploring inner mechanisms or layers. Conversely, we will use \textit{interpretability} for methods that define \textit{conceptual mapping mechanisms} whose goal is to contextualize models' predictions by associating them with an understandable concept, which in this paper is the syntax of programming languages. %

\textbf{Related Work on Interpretability in NLP.} There are existing techniques in both natural language processing (NLP) and SE literature focused on interpretability, including LIME~\cite{ribeiro2016should}, Kernel SHAP~\cite{lundberg2017unified}, Integrated Gradient~\cite{sundararajan2017axiomatic} and Contextual Decomposition~\cite{murdoch2018beyond}. These techniques generally try to approximate an interpretable model that either attempts to attribute meaning to hidden representations of neural networks, or illustrate the relationship between input features and model performance. However, we argue that such techniques are difficult to make practical in the context of LLMs for code, given the lack of conceptual mappings explained earlier. However, the most closely related interpretability technique to \approach, and one of the only to have adapted to LLMs of code is that of \textit{probing} which is a supervised analysis to determine which type of parameters (\eg input code snippets, tokenization process, number of hidden layers, and model size) influence the learning process in ML models \cite{troshin_probing_2022}. Probing aims to assess whether hidden representations of \llms encode specific linguistic properties such as syntactic structures of programming languages. Given our generated visualizations, there may be an inclination to characterize \approach as a \textit{probing technique}.
However, it is important to note that \approach is focused on estimating the \textit{correctness} of predicted syntactic code elements rather than mapping meaning to internal model representations of data. %

\textbf{Related Work on Interpretability in SE.} In the realm of SE research, prior work has taken two major directions: (i) techniques for task-specific explanations~\cite{fu2023vul-explain,liu2022explainable,pornprasit2021explain}, and (ii) empirical interpretability studies using existing NLP techniques~\cite{liu2024reliability,tantithamthavorn2023explain,mohammadkhani2023explain}. Previous authors have proposed techniques for explaining specific tasks including vulnerability explanation~\cite{fu2023vul-explain}, vulnerability prediction for Android~\cite{liu2022explainable}, and defect prediction models~\cite{pornprasit2021explain}. More recently Liu \etal conducted large empirical study using existing explainability techniques for global explanations of code to better understand generative language models of code~\cite{liu2024reliability}. Mohammadkhani \etal conducted a study using LLM's attention mechanism to interpret their performance on generating code. Finally, one paper that proposed a code-specific interpretability technique is that of Cito \etal \cite{cito2022counterfactual} who formulated a method to generate explanations using counterfactual reasoning of models. Our work on \approach complements this body of past work by developing a \textit{new, generally applicable interpretability method} that can be applied to \textit{both} local and global explanations of code, which no prior study or technique has done.

\textbf{Trustworthiness.} This research is inspired by definitions of \textit{trust} from automated systems, SE, and NLP. In automated systems, trust is defined as \textit{``the attitude that an agent will help achieve an individual's goal in a situation characterized by uncertainty and vulnerability''}~\cite{Lee_See_2004}. Bianco \etal define software trust as the degree of confidence when the software meets certain requirements~\cite{softtrust}. In NLP, Sun \etal argue that \llms must appropriately reflect truthfulness, safety, fairness, robustness, privacy, machine ethics, transparency, and accountability for them to be trustworthy~\cite{sun2024trustllm}. We define trust as the confidence that practitioners and researchers have in \llms' code prediction, anticipating that these predictions will effectively align with their intended goals. Trustworthiness in \llms implies a sense of interpretability in a given \llm's performance, instilling confidence among practitioners in their abilities to perform code-related tasks. To the best of our knowledge, no paper proposes a concrete definition of trust based on interpretability within the SE research community. Yet, several researchers have called for the importance of trustworthiness in \llms for code~\cite{lo_trustworthy_2023,spiess2024quality}. In our work we present a concrete definition of trustworthiness, highlight its importance, and show how syntax-grounded explanations such as \approach contribute to more trustworthy \llms.

\section{Syntax-Grounded Explanations}\label{sec:approach}

At a high level, \approach queries a \llm for probabilities per token, estimates the median across tokens that are part of one AST node, and presents those averages as \textbf{confidence performance} values segregated by hand-assigned syntax categories. We also refer to this confidence performance as \textit{\approach Interpretability Performance}.

\approach consists of four steps depicted in \figref{fig:overview}. In step \circled{1}, a code snippet for local or a testbed for global explanations is the starting point of the \textit{interpretability process}. Each sequence within the snippet or the testbed is processed by a tokenizer (\eg Byte-Pair Encoding (BPE)). In step \circled{2}, the tokenizer sets a vocabulary we named \textbf{token set}. Once code sequences are preprocessed, an \llm under analysis generates \textbf{token-level predictions} (\ntp) for each position in a sequence. Next, in step \circled{3}, the generated token-level predictions are aligned with the associated Abstract Syntax Tree (AST) terminal nodes. Terminal nodes only store  \ntp, while non-terminal nodes hierarchically store clustered and aggregated \ntp. Terminal and non-terminal nodes comprise the \textbf{subcategory set}. For example, consider \fbox{\texttt{\small if\_}} BPE token from the \textit{token set}. This token is aligned with the {\texttt{\small `if'}} terminal AST node while clustered in the \texttt{\small `if\_statement'} non-terminal node. Finally, in step \circled{4}, ten \textit{syntax categories} are proposed to summarize a model's predictions. \textit{Syntax Categories} aim to group the sub-categories into higher-level, more human-understandable categories. These syntax categories are a fixed \textbf{category set} that comprises more interpretable elements and include:
\begin{multicols}{3}
    \begin{itemize}
        \item {\small Decisions}
        \item {\small Data Structures}
        \item {\small Exceptions}
        \item {\small Iterations}
        \item {\small Functional Programming}
        \item {\small Operators}
        \item {\small Testing}
        \item {\small Scope}
        \item {\small Data Types}
        \item {\small Natural \\Language}
    \end{itemize}
\end{multicols}

For instance, the sub-categories \texttt{\small `if\_statement'} and  \texttt{\small `if'} are both clustered into one syntax category \textit{Decisions}. In the end, \approach generates an averaged score per category for global explanations and an AST tree visualization with stored scores at each node for local explanations. In essence, we propose that \textit{syntax elements} contain semantic information that contextualizes predicted probabilities. However, this semantic information varies across the granularity of these elements. We can claim, for example, that token-level elements carry less interpretable information than category-level elements.

\approach produces \textit{post-hoc} local and global explanations of generated code snippets. A local explanation intends to interpret the generation of a code snippet by decomposing it into AST elements. Conversely, a global explanation uses a set of generated snippets (or existing benchmark dataset) to interpret a given model holistically into \textit{Syntax Categories} (\ascs). The following sub-sections introduce the building blocks of syntax-grounded explanations. \secref{sec:composable} defines the interpretable sets (\eg \textit{Token}, \textit{Subcategory}, and \textit{Category}) that contain the \textit{syntax elements} employed for the interpretability process. \secref{sec:formalism} formalizes two function interactions that communicate previously interpretable sets. Such communication consists of aligning and clustering elements from code tokens to syntax categories. Finally, \secref{sec:explainations} shows the process of generating local and global explanations.

\subsection{Interpretable Syntax Sets}\label{sec:composable}
\textbf{Token Set $\mathcal{V}$.} Although \approach was designed to be compatible with different types of \llms, this paper concentrated on \textit{Decoder-Only} models due to their auto-regressive capacity to generate code~\cite{xu_systematic_2022} by preserving \textit{long-range dependencies}~\cite{karpathy2015understand}. A Decoder-only model can be employed as a generative process such as any token $w_i$ is being predicted by $\hat{w_i} \backsim P(w_i | w_{<i} ) = \sigma(y)_i = e^{y_{w_i}} / \Sigma_j e^{y_j}$. The term $y_j$ represents the \textit{non-normalized log-probabilities} for each output token $j$ (see \figref{fig:local_explaination}). We extracted and normalized these log-probabilities from the last layer of \llms to estimate \textit{Token-Level Predictions (\ntp)}. This estimation relies on the softmax function. The softmax $\sigma_i$ returns a distribution over predicted output classes, in this case, the classes are each token in the \textit{token set} $\mathcal{V}$. The predictions $\sigma_i$ are expected to be influenced by previous sequence inputs $w_{<i}$.

\textbf{Subcategory Set $\mathcal{N}$.} This set comprises elements of Context-Free Grammars (CFGs). Such elements are rules containing the syntax and structural information of a programming language~\cite{10.5555/1196416}. CFGs define instructions that specify how different tokens (\ie Lexemes) are assembled to form valid statements for each language. Formally, a $CFG$ is defined as $\mathbb{G} = (\alpha, \lambda, \omega, \beta)$ where $\alpha$ denotes the finite set of non-terminal nodes, $\lambda$ the finite set of terminal nodes, $\omega$ the finite set of production rules and $\beta$ the start symbol. CFGs use the terminal and non-terminal nodes (or subcategories) to define the production rules $\omega$ for any statement (\eg conditional, assignation, operator). Furthermore, these terminal and non-terminal nodes retain different meanings. Note that these nodes are the elements of the subcategory set $\lambda,\alpha \in \mathcal{N}$.

\textbf{Category Set $\mathcal{C}$.} The steps three and four in \figref{fig:overview} illustrate the binding of $\alpha$ and $\lambda$ into a category $c \in \mathcal{C}$. We pose the term \textbf{Syntax Categories} (\ascs) as the elements within the Category Set $\mathcal{C}$. We propose ten different \ascs based on tree-sitter bindings~\cite{tree-sitter} for Python. \ascs are the semantic units to enable the syntax interpretability of \llms. As such, \approach allows for Token-Level Predictions (\ntp) to be explained in a developer-centric way. In summary, each token in a sequence $s$ can be mapped to a category $c \in \mathcal{C}$. With \approach, practitioners can easily associate \llms' code predictions to specific structural attributes. For instance, \texttt{\small `identifier'} and \texttt{\small `string'} nodes correspond to a common \textit{Natural Language} category in \figref{fig:global_explanation}. As such, we can group nodes $\lambda$ and $\alpha$ into semantically meaningful \textit{categories} $\mathcal{C}$.

\subsection{Alignment and Clustering Formalism}\label{sec:formalism}

\begin{figure}
		
		\includegraphics[width=\textwidth]{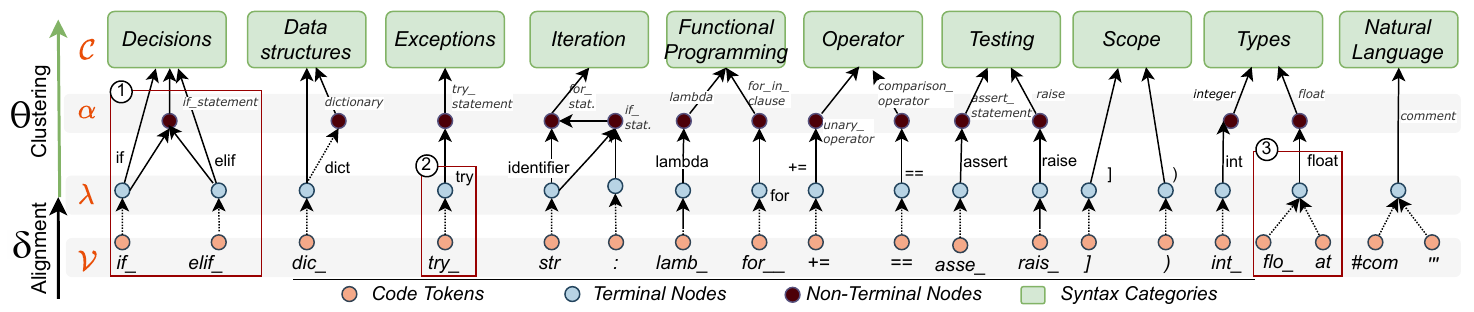}
		\caption{{\textit{Alignment \& Clustering Interactions}. The $\delta$ function aligns tokens $w_i$ to terminal nodes $\lambda$. Terminal and Non-terminal nodes $\lambda$, $\alpha$ $\in \mathcal{N}$ are clustered by Syntax Categories $\mathcal{C}$.}}
        \label{fig:runner_example}
 
\end{figure}

The previous subsection describes the syntax elements for enabling \llms interpretability (\ie \textit{token-set}, $\alpha$ and $\lambda$ subcategories, and \textit{Syntax Categories} (\ascs)). This section elaborates on the interaction among these elements. Two interactions in the form of a function are defined. { First, the alignment function $\delta$ links code tokens  from the \textit{Token Set}  $\mathcal{V}$ to terminal nodes $\lambda$. Second, the clustering function $\theta$ groups the subcategories $\lambda$ (terminal nodes) and $\alpha$ (non-terminal nodes) by syntax categories (\ascs) from the \textit{Category Set}. \figref{fig:runner_example} showcases both function interactions $\delta$ and $\theta$ respectively.}  

{\textbf{Alignment Interaction.} \figref{fig:runner_example} illustrates the process of aligning the terminal nodes $\delta$ in the AST to their corresponding code tokens $w_i$. This alignment starts by decomposing an input snippet $s$ into tokens $w_{<=i}\in \mathcal{V}$. For instance, \figref{fig:runner_example}-\circled{2}  depicts the alignment of  \fbox{\texttt{\small try\_}} token to the terminal $\lambda$ \texttt{\small `try'} node. Note that the alignment ignores the character "\_" from \fbox{\texttt{\small try\_}}. A tokenizer may produce a sequence in which each token does not necessarily match one-to-one with a terminal $\lambda$ node, \eg \figref{fig:runner_example}-\circled{3} illustrates the tokens \fbox{\texttt{\small flo\_}} and \fbox{\texttt{\small at}} are aligned with the $\lambda$ node \texttt{\small `float'}. Formally, $\delta(flo\_,at) \to [float]$ in a many-to-one interaction. Consequently, the alignment between code tokens and terminal nodes is certainly many-to-one, including one-to-one, but never one-to-many or many-to-many. 
}

\begin{definition}
\label{def:delta}
\textit{Alignment ($\delta$).} The function $\delta: w_{<=i} \to \vec{\lambda}$ where $w_{<=i}$ corresponds to a code sub-sequence whose tokens are many-to-one associated to the corresponding terminal node vector $\vec{\lambda}$ of syntax subcategories.

\end{definition}
{
\textbf{Clustering Interaction.} A clustering function $\theta$ estimates the \textbf{confidence performance} of $\lambda$ and $\alpha$ nodes (subcategories) from an AST by hierarchically aggregating the Token-Level Predictions (\ntp) to a \textit{Category} $c\in\mathcal{C}$. Once the tokens are aligned with their corresponding nodes using $\delta$ from Def.\ref{def:delta}, \approach clusters them into their respective category or non-terminal $\alpha$ node according to the AST representation. Some terminal $\lambda$ nodes can directly be aggregated into a category without considering intermediate non-terminal $\alpha$ nodes. A terminal $\lambda$ node can initiate a block sentence (\ie a category) and a block sequence parameters (\ie non-terminal if\_statement node). For instance, \figref{fig:runner_example}-\circled{1} depicts the terminal $\lambda$ \texttt{\small `if'} node aggregated into the \textit{Decisions} category and also starts the non-terminal $\alpha$ \texttt{\small `if\_statement'} node. To estimate the confidence performance, we traverse the entire AST and aggregate the \ntp probabilities of respective tokens. 

The $\theta$ function can adopt average, median, or max aggregations depending on the user configuration. \figref{fig:local_explaination} shows the clustering function applied to a concrete code generation sample. This application constitutes a local post hoc explanation: the parent node \texttt{\small `parameters'} has a $0.23$ associated confidence performance. This parent node average was aggregated with its terminal values: \texttt{\small `('} with $0.07$, \texttt{\small `identifier'} with $0.4$ and $0.1$, \texttt{\small `,'} with $0.5$, and \texttt{\small `)'} with $0.1$. Formally, $\theta( \vec{\lambda} = [0.07, 0.4, 0.1, 0.5, 0.1] ) \to [(parameters, 0.23)]$. If a sample snippet does not contain any particular syntax element (\ie token, subcategory, or category), such an element \textit{is therefore never considered for clustering}. An absent syntax element is reported as a \texttt{null} value to avoid biased syntax-grounded explanations. 
}

\begin{definition}
\label{def:theta}
\textit{Clustering ($\theta$).} The function $\theta: \vec{\lambda} \to median(\vec{n})$ where $\vec{\lambda}$ is the resulting vector of a sub-sequence and $\vec{n}$ is the vector of hierarchical associated non-terminal nodes for each terminal $\lambda$. The vector $\vec{n}$, therefore, contains the \ntp of non-terminal and the corresponding terminal nodes\footnote{In our study, we set the aggregation $\theta: N \to median(\hat{w}_{<=i})$ for a subset of tokens $w_{<=i}$.}. %
\end{definition}

\subsection{Post Hoc Local and Global Explanations}\label{sec:explainations}

\llms are more understandable when they \textit{reflect human knowledge}~\cite{Kim2018InterpretabilityTCAV}. One way of determining whether an \llm trained on code reflects human knowledge is testing it to see whether or not it operates \textit{similar to how a developer would estimate the prediction of a sequence}~\cite{palacio2023theory}. \approach can adopt the form of a \textit{post-hoc} local or a global explanation to make code predictions humanly understandable.  

\begin{figure}[t]
		\centering
		\includegraphics[width=\textwidth]{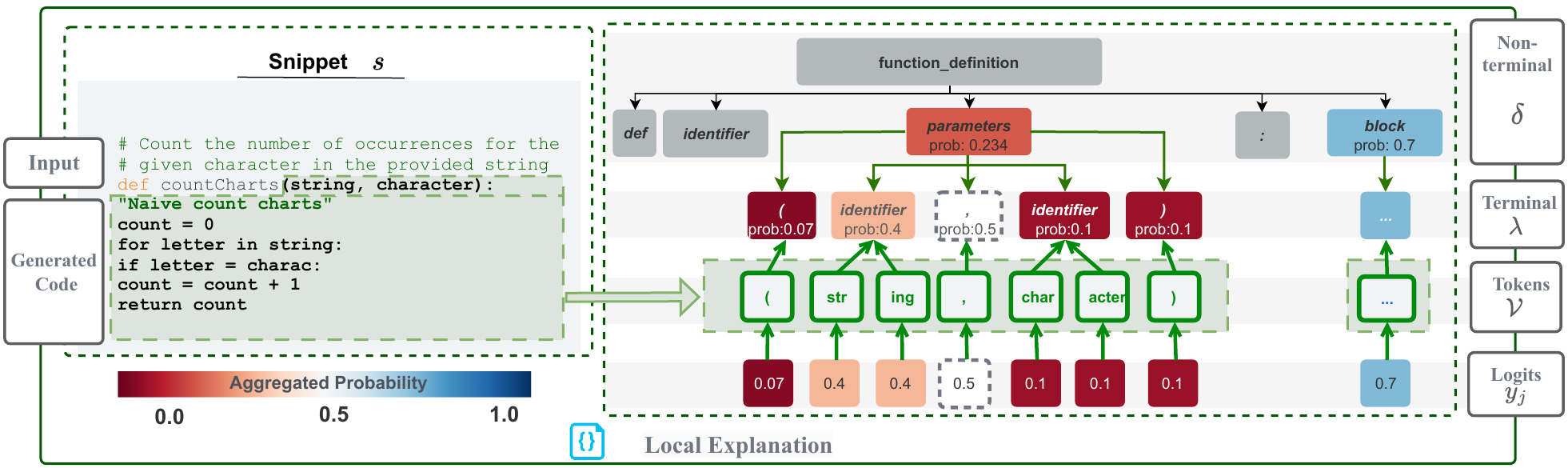}
		\caption{\textit{Post Hoc Local Explanation}.  A snippet is decomposed into code tokens. The highest annotated probabilities (\ie best predictions) are in blue.}
        \label{fig:local_explaination}
    \Description{Post hoc local explanation}
\end{figure}

\textbf{\approach for local interpretability} allows us to interpret a single snippet $s$ by generating a visual explanation based on an Abstract Syntax Tree (AST) as illustrated in \figref{fig:local_explaination}. A practitioner can explain the code predictions observing the probabilities associated with each element on the AST. In other words, we use $\theta$ from Def.~\ref{def:theta} to cluster around AST nodes across all levels (\ie AST probability annotations). Therefore, the syntax-grounded local explanation comprises a \textit{conceptual mapping} from the code prediction to a terminal and non-terminal node (or sub-categories). Fig.~\ref{fig:local_explaination} is a visual representation of the conceptual mapping using code predictions by \gptII model. The visualization displays a confidence value for each $\lambda$ and $\delta$ sub-categories after parsing the AST. The auto-completed snippet is processed with the clustering $\theta$ function.  

\begin{figure}[t]
\centering
\includegraphics[width=0.95\textwidth]{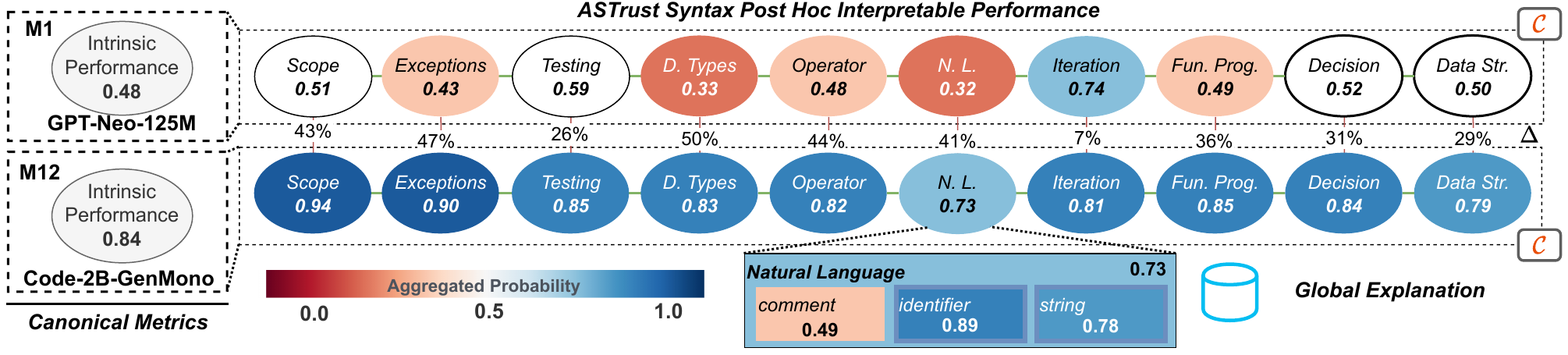}
\caption{\textit{Post Hoc Global Explanations} Segregated by Categories and Sub-Categories for \gptI and \monoIIII} 
\label{fig:global_explanation}
\Description{Global explanation post hoc explanation}
\end{figure}
{ 
\textbf{\approach for global interpretability} allows us to interpret a \llm by decomposing the canonical performance into segregated confidence performance. This segregated confidence is attached to Syntax Categories (\ascs). \ascs are tied to AST tree-sitter nodes~\cite{tree-sitter} and inspired by common object-oriented programming definitions. Although our approach is focused on the Python syntax, these categories may apply to multiple Programming Languages as they are being processed for similar AST elements.

Syntax-grounded global explanations comprise a \textit{conceptual mapping} from code predictions to ten syntax categories. These predictions are calculated for the entire testbed rather than a particular snippet following Def.~\ref{def:theta} about clustering function $\theta$. This clustering has an extra step in which a bootstrapping mechanism is used to estimate the \textit{confidence performance mean} of the category set.

Consequently, practitioners can compare Syntax Categories among models and explain the overall behavior of a given \llm by observing the segregated confidence performance. Note that previous analysis can be enriched with the information provided by canonical metrics. For instance, \figref{fig:global_explanation} depicts $M1$ with an intrinsic performance of $0.48$ while showing a \approach interpretability performance of $0.74$ for the \textit{iteration} category. Details of this global categorization can be further explored and visualized in our online appendix~\cite{AnonyRepoASTrust24}}.

{It should be noted that the validity of both global and local syntax-grounded explanations is dependent upon the \textit{calibration} of LLMs for code in terms of token probabilities and prediction correctness. That is, we assume that token probabilities are a reasonable proxy for the likelihood that a model is correct about a given token prediction. Recent work on calibration for LLMs of code has illustrated that, for code completion (which subsumes the experimental settings in this paper), LLMs tend to be well calibrated to token probabilities~\cite{spiess2024quality}. We also further confirm this finding using causal inference in Section~\ref{sec:global_validity}.}

\section{Empirical Study Design}\label{sec:emprical}

{We study the applicability of \approach in interpreting code completion tasks. We conducted a \textit{human study} to investigate the \textit{usefulness} of local explanations in real-world settings. In contrast, we conducted a \textit{data science study} to showcase the \textit{effectiveness} of global explanations on a diverse set of LLMs. Finally, we carried out a \textit{causal inference study} to assess the \textit{validity} of the syntax-grounded explanations as they relate to the statistical learning error of the studied models. The following research questions were formulated:}

\begin{enumerate}[label= \textbf{RQ$_{\arabic*}$}, ref=RQ$_{\arabic*}$, wide,labelindent=5pt]\setlength{\itemsep}{0.2em}
    \item \label{rq:survey_usefulness} \textbf{[Usefulness]} \textit{How useful are local explanations in real-world settings?} We validate the extent to which AST probability annotations are useful in locally explaining code predictions. We measure usefulness in three key factors: complexity, readability, and \llms' reliability.
    \item \label{rq:performance} \textbf{[Effectiveness]} \textit{To what extent do \llms for code correctly predict different syntactic structures?} We interpret the performance of 12 \llms on each Syntax Category (\asc). The conceptual mapping allows us to obtain an interpretable and segregated confidence value per category, so we can detect categories that are easier or harder to predict -- moving beyond canonical aggregate metrics. 
    \item \label{rq:validity} \textbf{[Validity]} \textit{How do Syntax Concepts impact LLMs' statistical learning error?} We validate the causal connection between learning error and \llms' ability to predict different syntax categories using \approach.
\end{enumerate}

\subsection{Experimental Context}

\subsubsection{Model Collection.} To perform our global analysis, we conducted an interpretability analysis of 12 open Decoder-only \llms, selected based on their popularity. The largest among these models boasts 2.7 billion parameters. \tabref{tab:models} categorizes these \llms into four distinct groups, each aligned with a specific fine-tuning strategy. The initial category comprises GPT-3-based models primarily trained on natural language, exemplified by Pile \cite{gao2020pile}. The second category encompasses models trained on natural language but constructed upon the \textit{codegen} architecture \cite{nijkamp2023codegen}. Moving to the third category, we find models trained on multiple programming languages (PLs) using BigQuery \cite{bigquery}, implemented on both the gpt-2 and codegen architectures. The final category consists of both \multilang models fine-tuned on BigPython \cite{nijkamp2023codegen}, denoted as \monolang, and gpt-2 models such as codeparrot \cite{codeparrot}. All the datasets for training the \llms encompass repositories/files sourced from GitHub up to 2021.

\subsubsection{Evaluation Dataset} To ensure the integrity of our \approach evaluation, it is imperative to avoid data contamination by excluding samples used in the training process of the \llms. We extended and used \galeras \cite{10336302} to overcome this challenge. \galeras exclusively comprises code commits from the top 200 most popular Python GitHub repositories between 01/01/22 and 01/01/23. Notably, \galeras incorporates comprehensive data, including commit messages, method comments, the entire AST structure, node count, AST levels, AST errors, whitespace details, lines of code, cyclomatic complexity, and token counts.

\subsubsection{Machine Configuration.} We performed the experiments using 20.04 Ubuntu with an AMD EPYC 7532 32-Core CPU, A100 NVIDIA GPU with 40GB VRAM, and 1TB RAM. For the model inference process, we used HugginFace and Pytorch \cite{wolf2020transformers, pytorch}. All models were loaded into the GPU to boost the inference time.

\begin{table}[t]
\centering
\caption{Large Language Models Descriptions.}
\vspace{-0.2cm}
\label{tab:models}

\scalebox{0.90}{%
\setlength{\tabcolsep}{5pt} 

\begin{tabular}{lllll}
\hline
\multicolumn{5}{c}{\textbf{Large Language Models (LLMs)}} \\ \hline
\multicolumn{1}{c}{\textit{Type}} & \textit{ID} & \multicolumn{1}{c}{\textit{Name}} & \multicolumn{1}{c}{\textit{Architecture}} & \multicolumn{1}{c}{\textit{Size}} \\ \hline
\multirow{3}{*}{\textbf{\begin{tabular}[c]{@{}l@{}}Natural L.\\ gpt-3\end{tabular}}} & $M_{1}$ & gpt-neo-125m & \textit{gpt-3} & 125M \\
 & $M_{2}$* & gpt-neo-1.3B & \textit{gpt-3} & 1.3B \\
 & $M_{3}$ & gpt-neo-2.7B & \textit{gpt-3} & 2.7B \\ \hline
\multirow{2}{*}{\textbf{\begin{tabular}[c]{@{}l@{}}Natural L.\\ codegen\end{tabular}}} & $M_{4}$ & codegen-350M-nl & \textit{codegen} & 350M \\
 & $M_{5}$ & codegen-2B-nl & \textit{codegen} & 2B \\ \hline
\multirow{3}{*}{\textbf{\begin{tabular}[c]{@{}l@{}}Multi-\\ Language\end{tabular}}} & $M_{6}$ & codeparrot-small-multi & \textit{gpt-2} & 110M \\
 & $M_{7}$ & codegen-350M-multi & \textit{codegen-350M-nl} & 350M \\
 & $M_{8}$ & codegen-2B-multi & \textit{codegen-2B-nl} & 2B \\ \hline
\multirow{4}{*}{\textbf{\begin{tabular}[c]{@{}l@{}}Mono-\\ Language\end{tabular}}} & $M_{9}$ & codeparrot-small & \textit{gpt-2} & 110M \\
 & $M_{10}$ & codeparrot & \textit{gpt-2} & 1.5B \\
 & $M_{11}$ & codegen-350M-mono & \textit{codegen-350M-multi} & 350M \\
 & $M_{12}$ & codegen-2B-mono & \textit{codegen-2B-multi} & 2B \\ \hline
\end{tabular}

}
\vspace{0.1cm}
{\\ \footnotesize{*The human study was conducted using $M_{2}$ - \gptII}}

\vspace{-0.5cm}

\end{table}

\subsection{Human Study for \approach Usefulness}
\label{sec:local_study_design}

{
This section presents a preliminary human study comprising a control/treatment experimental design to assess \approach usefulness in practical settings. We followed a \textbf{purposive sampling approach} \cite{baltes_sampling_2021} since our primary goal was to gather preliminary data and insights from practitioners with expertise in ML and SE combined. We selected our subjects carefully rather than randomly to study the usefulness of \approach (at local explanations). \approach is designed to enhance the interpretation of model decisions in code completion tasks for practitioners with diverse backgrounds, including researchers, students, and data scientists. By targeting individuals with specific expertise, we ensured that the feedback received was relevant and informed, thereby enhancing the quality of our preliminary findings.}

\subsubsection{Survey Structure} {
Each survey consists of three sections. The \textit{first section} is aimed at gathering participant profiling information. The profiling section aims to collect information related to how proficient the participants are when using Python and AI-assisted tools in code generation tasks. In particular, we asked about their level of expertise in Programming Languages (PL) and how familiar they are with AST structure. Furthermore, we asked about any challenges they encountered when using AI-assisted tools. This information is relevant because we want to control external factors that may influence their perception in validating \approach. The \textit{second section} aimed to present four code completion scenarios and ask participants to rate their quality. For each scenario, we presented an incomplete Python method as a prompt, the generated code completing the previous method (using \gptII in \tabref{tab:models}), and a specific type of local explanation (\eg \surveySequence, \surveyAstPartial, and \surveyAstComplete in \figref{fig:sec_8_survey_explanations}). {In all four scenarios, the prompt contained either a \textit{semantic error} (\eg using an undefined variable, incorrect condition statement(s), calling an undefined function) or a 
\textit{syntax error} (\eg missing colon, missing parenthesis, incorrect indentation) that participants needed to reason about after considering a given local explanation type. We aimed to capture the participant's perspective regarding the explanation by asking them to describe the cause of a syntax or semantic error from the generated code. To facilitate the analysis, we highlighted the portion of the code generated by the model (see green highlight \figref{fig:local_explaination}). We did not provide any detail about the \llm used to generate the predictions in order prevent introducing bias related to preconceived notions about particular LLMs in the responses. Lastly, the \textit{third section} aim to collect feedback about the effectiveness of the different types of local explanation. Specifically, we asked participants about the complexity of the visualizations and potential opportunities for enhancement.}

\subsubsection{Survey Treatments}

To collect the perception of practitioners regarding the usability of \approach, we devised a control survey (\surveyControl) and three treatments with two types of local explanations: sequential (\surveySequence) and AST-based (\surveyAstBased) explanations. \surveyControl represents \textit{the absence of a local explanation} and only collects the participants' perceptions regarding the \textit{correctness} of \llms' output. By contrast, treatment surveys \surveySequence and \surveyAstBased yield syntax-grounded explanations. It is worth noting that {we define \textit{correctness} as the degree to which the generated code reflects the purpose of the algorithm contained in the prompt. In other words, we ask participants to judge whether the model predicted a valid or closely accurate set of tokens given the information context within the prompt.}

{ \figref{fig:sec_8_survey_explanations} depicts all the explanation types considered in the survey. \surveySequence displays the tokens and their corresponding probabilities in a sequential (\ie linear layout). Linear representations are commonly used by feature-importance explainability techniques such as attention-based \cite{what_do_they_learn} and Shapley \cite{Kalai1983OnWS} values. Therefore, \surveySequence serves as a \textit{baseline} to determine how our local explanations \surveyAstBased provide insightful information beyond the linear layout. Conversely, \surveyAstBased uses an AST visualization, comprising two types of local explanations: AST-Complete (\ie \surveyAstComplete) and AST-partial (\ie \surveyAstPartial). \surveyAstComplete represents the entire sample's AST (\ie prompt and generated code) including \approach confidence performance for all nodes. Conversely, \surveyAstPartial is a filtered \surveyAstComplete representation that only exposes the confidence performance of the generated code and omits the nodes from the prompt.}

\subsubsection{Survey Metrics} When evaluating the usefulness of our approach to answer \ref{rq:survey_usefulness}, we measure the qualitative features of local explanations depicted in \figref{fig:sec_8_survey_explanations}. {More precisely, we proposed five qualitative metrics to evaluate the usefulness of our approach: \textit{Information Usefulness}, \textit{Local Explanation Complexity}, \textit{Local Explanation Readability}, \textit{Visualization Usefulness}, and \textit{LLM’s reliability}.} {We used a Likert scale with three options for quantitatively measuring the responses. Specifically for Information Usefulness: \textit{Agree}, \textit{Neutral} and \textit{Disagree}. For Local Explanation Complexity, Local Explanation Readability and Visualization Usefulness: \textit{Useful}, \textit{Slightly useful} and \textit{Not useful}. Finally, for LLM's Reliability: \textit{Not reliable}, \textit{Highly reliable} and \textit{Impossible to tell}. Each of the survey metrics corresponds to one of the following survey questions.}

\textit{{Metric$_1$: Information Usefulness} - `Q: How useful was the information for interpreting the model's decisions?'} In the treatment surveys, we ask the participants to explain the \llm's behavior when completing the code of individual samples, and we gauge their perception regarding the usefulness of the provided information to accomplish this task. We anticipate correlations between the explanation types and perceived usefulness.

\textit{{Metric$_2$: Local Explanation Complexity} - `Q: I found the visualization unnecessarily complex'}. The local explanation complexity refers to the degree of intricacy of its types. The degree of complexity may affect perceptions of usefulness.

\textit{{Metric$_3$: Local Explanation Readability} - `Q: I thought the visualization was easy to read and use'}. We define readability as the degree to which our local explanations are intuitive and easy to understand. We hypothesize that if the explanation fits this criterion, we can consider it useful. Readability accounts for factors such as the amount of consigned information, the arrangement of tokens and categories, and the color scheme.

\textit{{Metric$_4$: Visualization Usefulness} - `Q: I thought the visualization was useful for explaining the model's behavior'}. The visualization is the graphical representation of the local explanation (refer to \figref{fig:local_explaination}). Each treatment survey uses a type of visualization (\ie \surveySequence or \surveyAstBased). We formulate this question to determine which visualization is considered more useful.

\textit{{Metric$_5$: \llm's Reliability} - `Q: What is your perception of the model's reliability in generating code?'}. We define reliability as the degree to which a user trusts the \llm's output based on the outcomes from local explanations. We ask the participants to reflect on the \llm's reliability across our surveys using local explanations. Considering all four code completion scenarios in the surveys include errors, the greater the number of participants in each survey who would not rely on the model, the more valuable the syntax-grounded local explanation.

\subsubsection{Open Questions} 

{In addition to survey metrics, we formulated several open-ended questions for collecting the participants' perception about the correctness of the predictions (Open$_1$) and the most helpful parts of the visual explanations including potential improvement aspects (Open$_2$). Each of these open metrics corresponds to one or more survey questions. }

{\textit{{Open$_1$: \llm's Prediction Correctness} - `Q: If the generated code is incorrect, can you explain why the model might have made the mistake? Otherwise, If the generated code is correct, can you speculate on why the model may have been able to correctly predict the above snippet?'}. We asked the participants to use the provided information per sample to analyze whether the prompt or the generated code contained any syntax or semantic error. In \surveyControl, we aimed to assess the extent to which participants could reason about the source code correctness without any type of explanation provided. Conversely, in \surveySequence and \surveyAstBased, we inspected if the layout information somehow contributed to detecting and reasoning about the cause of the error.}

{
\textit{{Open$_2$: Importance of visual explanations} - `$Q_1:$ What information from the visualization did you find useful in explaining the model's predictions?', `$Q_2:$ What information from the visualization did you find useful in explaining the model's predictions?', `$Q_3:$ What other information (if any) would you like to see in the visualization?', `$Q_4:$ What elements of the visualization did you like most?', `$Q_5:$ What elements of the visualization did you like least?'}. We asked the participants to provide overall feedback about the type of representation used in the treatment surveys (\surveySequence and \surveyAstBased). We aimed to identify the most and least useful elements, as well as gather potential ideas for improvement.}

{To collect, standardize, and analyze the previous group of open-ended questions, two authors independently gathered and reviewed each survey's responses. Any differences were resolved through discussion to reach a consensus.}

\subsubsection{Population Profiling}
{The target population consists of software engineering practitioners experienced in using AI tools for code generation (\eg ChatGPT, Copilot). Participants were meant to be knowledgeable in Python and understand how algorithms are structured in programming languages and represented in Abstract Syntax Trees (ASTs). While certain knowledge in Deep Learning architectures used for Text Generation (\eg GPT, BERT, T5) is preferred, it was not required. Individuals of any gender were welcome to participate, with a minimum age requirement of 21 years. Participation was entirely voluntary, and no incentives were offered beyond contributing to our efforts to enhance deep learning interpretability for code generation. Furthermore, participants were informed of the voluntary nature of the study during solicitation \footnote{This study was reviewed by the protection of human subjects committee at the College of William \& Mary under protocol number PHSC-2023-03-03-16218-dposhyvanyk titled \textit{A Survey Research on Code Concepts for Interpreting Neural Language Models}}.}

\subsubsection{Data Collection}
{We reached out to $50$ potential participants who were unaware of the purpose of this work, from industrial and academic backgrounds with varying levels of expertise in machine learning and Python. Participants were contacted via email invitations. Out of this group, $27$ completed one of the surveys, with the assignment uniformly distributed among the surveys. but we excluded three for low-quality responses, leaving $24$ valid submissions. The study was performed on Qualtrics \cite{noauthor_qualtrics_nodate} and the anonymized survey data can be found in our appendix~\cite{AnonyRepoASTrust24}.}

\subsubsection{Statistical Analysis}
{We use \surveySequence as a baseline for our study. We expose the participants to \approach with two treatments: \surveyAstPartial and \surveyAstComplete (refer to \figref{fig:sec_8_survey_explanations}). The result of each question is influenced by these two treatments. To compare the influence of \surveyAstPartial and \surveyAstComplete against \surveySequence, we compute the weighted average of the responses from surveys \surveyAstPartial and \surveyAstComplete. We refer to the weighted average as \surveyAstBased. First, we calculate the results of each treatment individually for all the answers. Then, the weight of each answer is estimated by averaging the number of responses per answer across all samples. We then normalize this weight to get the final weighted average for \surveyAstBased. We use this weighted average for all our statistical analyses in the paper.}

\begin{figure}[t]
		\centering
        \vspace{-1em}
		\includegraphics[width=0.9\textwidth]{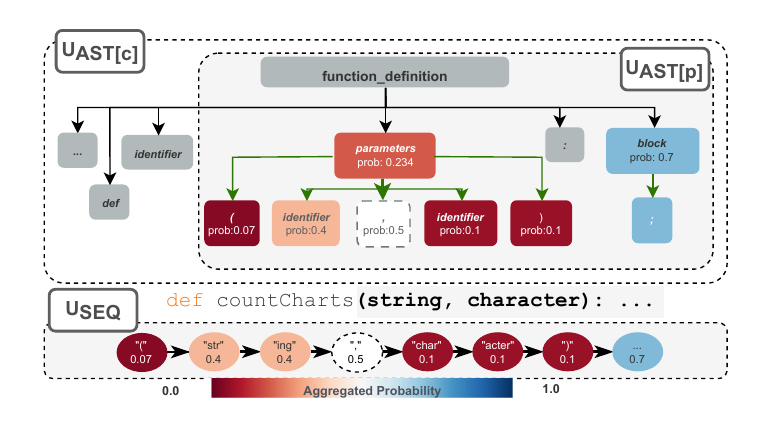}
        \vspace{-1.5em}
		\caption{\approach Local Explanation Treatments.}
        \vspace{-1.5em}
        \label{fig:sec_8_survey_explanations}
\end{figure}

\subsubsection{Survey Validity}
{To validate the design of the human study, we conducted a pilot experiment with 10 individuals excluded from the pool of participants. Based on this pilot, the quality and appropriateness of the control and treatment surveys were solidified.  Initially, the pilot survey included only the \surveyControl control and the \surveyAstComplete treatment. However, the pilot revealed the need for an intermediate representation serving as a baseline explanation, which is less complex than an AST visualization, to ensure a fair comparison. Consequently, we introduced \surveySequence, inspired by techniques such as SHAP \cite{lundberg2017unified}, as a baseline treatment with a less complex representation. Additionally, we introduced \surveyAstPartial, a partial representation of \surveyAstComplete, as a less complex treatment highlighting only the hierarchical structure of the generated code.}

\subsection{Data Science Study for \approach  Effectiveness}
\label{sec:global_study_design}
To answer \ref{rq:performance} we implemented a {data science study to globally} interpret 12 \llms' performance described in \tabref{tab:models} on the \galeras dataset. We performed code completion with different input prompts. The input prompt combines the code completion task, a description, and a partial code snippet. Each prompt has a standard maximum size of 1024 tokens for all considered \llms.

We first compute the normalized log-probabilities (Sec.\ref{sec:composable}) or \ntp $\hat{w_i}$ for each \galeras snippet $s \in \mathcal{S}$. These log-probabilities were obtained across the 12 \llms for every token position. The log-probability distributions maintain a consistent vector size $|\mathcal{V}|$ for each token position. Subsequently, these distributions underwent processing to extract the log-probability aligned with the expected token at position $i$. As a result, each token position corresponds to a stored prediction value $\hat{w_i}$ for constructing the \ntp sequence $w_{<=i}$. {As discussed earlier, this experimental setting is based on the premise that token probabilities are well-calibrated to model correctness, which has been confirmed in code completion settings by prior work~\cite{spiess2024quality}. Additionally, we confirm this finding in answering RQ$_3$ by observing a causal link between learning error and the probabilities used within \approach.}

We used the alignment function $\delta$ to obtain the terminal node $\lambda$ vector (see Def.\ref{def:delta}). Next, we traversed the AST for each terminal node $\lambda$ and clustered them into the corresponding final $\lambda,\alpha$ node and their correspondent \ntp by applying the $\theta$ function (see Sec.\ref{sec:approach}). The clustering was fixed to generate 32 subcategories and their probability values. We estimated a single \textbf{confidence performance} metric (\aka \approach Interpretability Performance) per model by averaging the subcategories probabilities. The confidence performance per model was bootstrapped with the median (size of 500 samplings) to ensure a fair comparison. Lastly, we mapped the subcategories to the \ascs obtaining a value per Category $\mathcal{C}$ (e.g., Data Structures, Decision, or Scope).

{To provide a baseline comparison, we calculated canonical extrinsic metrics \textit{BLUE-4}~\cite{papineni_bleu_2002} and \textit{CodeBLEU}~\cite{Ren2020codebleu}, and intrinsic performance.
Extrinsic metrics evaluate downstream tasks directly (\ie code completion), while intrinsic metrics assess how well a language model can accurately predict the next word given an incomplete sequence or prompt \cite{Karampatsis2020BigCode}.}

\begin{wrapfigure}{r}{0.25\textwidth}
  \begin{center}
    \includegraphics[width=0.25\textwidth]{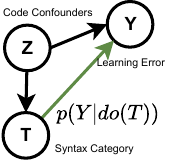}
  \end{center}
  \caption{\scm to estimate Syntax Effect on Learning Error.}
   \label{fig:causal_diagram}
   \vspace{-1em}
\end{wrapfigure}

Our analysis also includes a corner case experiment that compares the smaller \gptI to the largest \monoIIII. We contrasted the subcategories for each \llm to obtain a segregated global explanation \figref{fig:largeTreeMap}. Since we mapped the subcategories to categories, we can observe the \approach probability gaps between \llms at more interpretable levels (see. \figref{fig:global_explanation}). The probability values for subcategories and categories are the bootstrapped median.

\subsection{Causal Inference Study for \approach  Validity} We validate our \approach approach using causal inference to answer \ref{rq:validity}. To accomplish this, we formulated a \textbf{Structural Causal Model} (\scm) designed to estimate the impact of \asc predictions on the overall learning error of \llms \cite{Pearl2016Causality}. We consider that the learning error (\ie cross-entropy loss) of an \llm is causally impacted by the predicted probabilities of syntax elements. This impact indicates that \ascs influence the quality of an \llm. We conducted a causal inference analysis using the $do_{code}$ technique \cite{palacio2023theory} to estimate \ascs influence. Inherently, a developer mentally rationalizes several things such as the \textit{concept} of the \textit{Iteration} category (see Fig.~\ref{fig:largeTreeMap}). If an \llm can make a similar mapping, it suggests that it has \textit{statistically learned} some understanding of the syntax cycle structure.

We calculate the causal influence using the \scm as the Average Treatment Effect (ATE) with the probability $p(Y|do(T))$ for both \gptI and \monoIIII models.  That is, we estimate the \textit{causal effect} of the variable $T$ on $Y$ after controlling for confounders $Z$ (see \figref{fig:causal_diagram}). This probability function is estimated using the \textit{doWhy} tool ~\cite{Sharma2021DoWhyAssumptions, palacio2023theory}. The proposed treatments ($T$) embodies Syntax Categories $\mathcal{C}$ such as \textit{Decision}, \textit{Natural Language}, or \textit{Iterative}. 

The first step of this validity evaluation is to obtain the global intrinsic accuracy. We computed the cross-entropy loss for each snippet $s$. After obtaining the cross-entropy loss, we estimate Pearson correlation $\rho$ and ATE for 14 sub-categories ($\lambda$ and $\alpha$ nodes) (\tabref{tab:correlations}). Each sub-category and its cross-entropy loss is correlated with four confounding variables (\ie Cyclomatic Complexity, AST Levels, \#AST Nodes, and Sequence Size) calculating the average value from the set of snippets $S$ from \galeras dataset~\cite{10336302}. The second step is to validate the obtained ATE by testing the \scm robustness (\ie refutation methods ~\cite{palacio2023theory}). We limited our exploration to the best and worst models by intrinsic accuracy as we observed similar correlation values across LLMs.

\section{Results}\label{sec:results}

{In this section, we present our findings for human, data science, and causal studies. The local analysis is focused on answering \ref{rq:survey_usefulness} by using \approach to interpret concrete snippets. Similarly, we provide insights into our global analysis to answer \ref{rq:performance} and  \ref{rq:validity}, which incorporates the interpretation of \llms' performance segregated by Syntax Categories, a comparison of edge cases, and a causal assessment of \approach validity.

Before presenting the results, we point out basic stats about AST data processing: The average tree height of the samples in the empirical study was 30, with an average of 104 tokens and 166 nodes. In the human study, the four samples have distinct complexity levels. The smallest sample has 80 tokens, with 47 AST nodes and a tree of height eight. The biggest sample has a token length of 139, with 117 AST nodes and a tree height of 14.}

\subsection{\ref{rq:survey_usefulness} \approach Usefulness}
~\label{sec:local1}
Below, we present the results for each survey question as introduced in \secref{sec:local_study_design}. Quantified responses are detailed in \tabref{tab:user_study_quantitative_results}. In addition, we summarize the most relevant feedback received in the open-ended questions. The full human study's results can be accessed in the appendix~\cite{AnonyRepoASTrust24}.

\begin{table}[t]
\centering
\caption{Survey results for the \approach Local Study.} 
\label{tab:user_study_quantitative_results}
\vspace{-0.2cm}

\scalebox{0.75}{%
\setlength{\tabcolsep}{5pt} 
\begin{tabular}{cllccc}
\hline
\textbf{\textit{Survey Question}} &  & \multicolumn{4}{c}{\textbf{\textit{Results (\% answers)}}} \\ \hline
\textit{} &  &  & \textit{\textbf{Useful}} & \textit{\textbf{Slightly Useful}} & \textit{\textbf{Not useful}} \\
 &  & \textit{\surveySequence} & \cellcolor[HTML]{EFEFEF}\textbf{50.00} & \cellcolor[HTML]{EFEFEF}25.00 & 25.00 \\
 &  & \textit{\surveyAstBased} & 43.57 & \textbf{23.91} & \cellcolor[HTML]{EFEFEF}32.52 \\
 &  & \textit{\surveyAstPartial} & 39.29 & 28.57 & \textbf{32.14} \\
\multirow{-4}{*}{\textit{\begin{tabular}[c]{@{}c@{}}Information\\ Usefulness\end{tabular}}} &  & \textit{\surveyAstComplete} & 41.66 & \textbf{20.84} & \textbf{37.50} \\ \hline \hline
\textit{} &  &  & \textit{\textbf{Agree}} & \textit{\textbf{Neutral}} & \textit{\textbf{Disagree}} \\
 &  & \textit{\surveySequence} & 42.00 & \cellcolor[HTML]{EFEFEF}29.00 & \cellcolor[HTML]{EFEFEF}29.00 \\
 &  & \textit{\surveyAstBased} & \cellcolor[HTML]{EFEFEF}44.00 & \textbf{28.00} & \textbf{28.00} \\
 &  & \textit{\surveyAstPartial} & 14.00 & \textbf{43.00} & \textbf{43.00}\\
\multirow{-4}{*}{\textit{\begin{tabular}[c]{@{}c@{}}Local Explanation\\ Complexity\end{tabular}}} &  & \textit{\surveyAstComplete} &  \textbf{67.00} & 17.00 & 16.00\\ \hline
 &  & \textit{\surveySequence} & 29.00 & \cellcolor[HTML]{EFEFEF}29.00 & 42.00 \\
 &  & \textit{\surveyAstBased} & \cellcolor[HTML]{EFEFEF}\textbf{35.00} & 21.00 & \cellcolor[HTML]{EFEFEF}44.00 \\
 &  & \textit{\surveyAstPartial} & \textbf{57.00} & 29.00 & 14.00 \\
\multirow{-4}{*}{\textit{\begin{tabular}[c]{@{}c@{}}Local Explanation\\ Readability\end{tabular}}} &  & \textit{\surveyAstComplete} & 17.00 & 33.00 & \textbf{50.00} \\ \hline
 &  & \textit{\surveySequence} & \cellcolor[HTML]{EFEFEF}\textbf{57.00} & \cellcolor[HTML]{EFEFEF}29.00 & 14.00 \\
 &  & \textit{\surveyAstBased} & \textbf{49.80} & 27.78 & \cellcolor[HTML]{EFEFEF}22.42 \\
 &  & \textit{\surveyAstPartial} & \textbf{42.00} & 29.00 & 29.00 \\
\multirow{-4}{*}{\textit{\begin{tabular}[c]{@{}c@{}}Visualization\\ Usefulness\end{tabular}}} &  & \textit{\surveyAstComplete} & \textbf{50.00} & 33.00 & 17.00 \\ \hline \hline
 &  &  & \textit{\textbf{Highly Reliable}} & \textit{\textbf{Not Reliable}} & \textit{\textbf{Impossible to Tell}} \\
 &  & \textit{\surveySequence} & \cellcolor[HTML]{EFEFEF}29.00 & \textbf{42.00} & 29.00 \\
 &  & \textit{\surveyAstBased} & 0.00 & \cellcolor[HTML]{EFEFEF}\textbf{62.00} & \cellcolor[HTML]{EFEFEF}38.00 \\
 &  & \textit{\surveyAstPartial} & 0.00 & \textbf{57.00} & 43.00 \\
\multirow{-4}{*}{\textit{\begin{tabular}[c]{@{}c@{}}\llm's \\ Reliability\end{tabular}}} &  & \textit{\surveyAstComplete} & 0.00 & \textbf{67.00} & 33.00 \\ \hline
\end{tabular}
}
\vspace{0.1cm}
{\\ \footnotesize{* \textbf{bold}:Highest \%, background:Highest \% \surveySequence or \surveyAstBased}}
\vspace{-0.5cm}
\end{table}

\textit{Metric$_1$: Information Usefulness}. The data reveals that $67.48\%$ of participants who evaluated \surveyAstBased explanations, found the presented information useful or slightly useful, with a slight preference for \surveyAstPartial ($67.86\%$) over \surveyAstComplete ($62.5\%$). However, $75\%$ of participants who evaluated \surveySequence felt that it was useful, indicating a stronger preference towards it.

\textit{Metric$_2$: Local Explanation Complexity}. Participants found \surveyAstBased explanations slightly more complex ($44\%$) than \surveySequence ($42\%$). In particular, \surveyAstComplete was found substantially more complex ($67\%$) than \surveyAstPartial. This is not surprising, given that complete ASTs, even for small code snippets can appear complex.

\textit{Metric$_3$: Local Explanation Readability}. Both \surveyAstBased and \surveySequence were found to be similarly readable: $35\%$ participants found \surveyAstBased easy to read and use, compared to $29\%$ for \surveySequence. However, between the two AST types \surveyAstPartial ($57\%$) was far preferred in contrast to \surveyAstComplete ($17\%$), again likely due to the complexity of \surveyAstComplete.

\textit{Metric$_4$: Visualization Usefulness}. \surveySequence visualization was found useful by more than half of the participants who evaluated it ($57\%$). Similarly, $49.8\%$ considered the \surveyAstBased visualizations useful, with an appreciable preference for  \surveyAstComplete ($50\%$) over \surveyAstPartial ($42\%$).

\textit{Metric$_5$: \llm's Reliability}. The high number of participants who judged the \llm as unreliable suggests that all types of explanations helped them assess the quality of the predicted code. However, \surveyAstComplete stood out, with $67\%$ participants favoring it. Meanwhile, $29\%$ participants felt confident about the model based on the information presented by \surveySequence, indicating the potential of formulating incorrect interpretations when using this type of explanation. Interestingly, a high percentage of participants ($43\%$) found that \surveyAstPartial cannot help to conclude whether \llm is reliable.

{\textit{Open$_1$: \llm's Prediction Correctness.} Participants attributed the cause of an incorrect prediction in the model's output to syntax and semantic errors in both treatment and control surveys. The attribution to training bias was prevalent in \surveyControl as evidenced in answers such as \textit{``Model has not seen enough samples to differentiate between [the characters] = and ==''} or \textit{``Maybe the model is trained in problems with similar error''}. However, in \surveySequence and \surveyAstBased those responses included attribution to the low \approach confidence performance, such as \textit{``The probabilities are very low so the predictions are not correct''} and \textit{``[the character] = has low probability score of 0.0096''}. These results reveal that \approach explanations provided insightful information for the participants to judge the model's decisions. }

{\textit{Open$_2$: Importance of visual explanations.} Participants favored the color scheme and the \approach confidence performance associated with each token as the most liked elements in the visual explanations. Conversely, they disfavored the inclusion of certain syntax-related tokens, such as white spaces and punctuation marks, in the interpretability analysis. We also encountered contradictory premises: \surveyAstPartial participants believe the explanation missed important details, while in \surveyAstComplete participants criticized the information overload. Participants also suggested improving the navigation in \surveyAstBased representations by incorporating a mechanism to interactively collapse AST nodes.}

\textit{Profiling.} We found that the participants were well-qualified to take our survey. They all had some background in Machine Learning (Formal or Informal). Similarly, 81.25\% of participants were also familiar with the concept of AST.

\begin{boxK}
\ref{rq:survey_usefulness}: Although Sequence Explanations contain useful information, AST visualizations were viewed most favorably among explanation types. In fact, AST-based Explanations were found most effective to judge \llm reliability.
\end{boxK}

\subsection{\ref{rq:performance} \approach Effectiveness}
~\label{sec:global_performance}

{To answer \ref{rq:performance} we computed both the canonical intrinsic performance and the \approach interpretability performance for 12 \llms (\tabref{tab:models}). \figref{fig:canonical_performance} depicts the canonical intrinsic performance for each \llm (\ie box-plot) and the density canonical intrinsic performance (\ie density plot) by model type (\eg \nlgpt, \nlcodegen, \monolang, and \multilang). The intrinsic performance comprises an aggregated metric that allows us to compare models at a glance. For instance, on average the smallest \monoI ($M_9$) has a similar intrinsic performance as the largest GPT-based \gptIII ($M_3$) model with intrinsic performance of $0.61$ and $0.62$ respectively. After grouping the models by types, we observe that \monolang models excel in the intrinsic performance with the highest density of $0.9$ for performance values between $(0.6 - 0.8)$ and an average intrinsic performance of $\approx 0.7$. Despite the fact canonical intrinsic performance can statistically describe, on average, how the model performs at generating code, these metrics are limited to explaining which categories are being predicted more confidently than others.}

\begin{figure}[ht]
\centering
\includegraphics[width=0.95\textwidth]{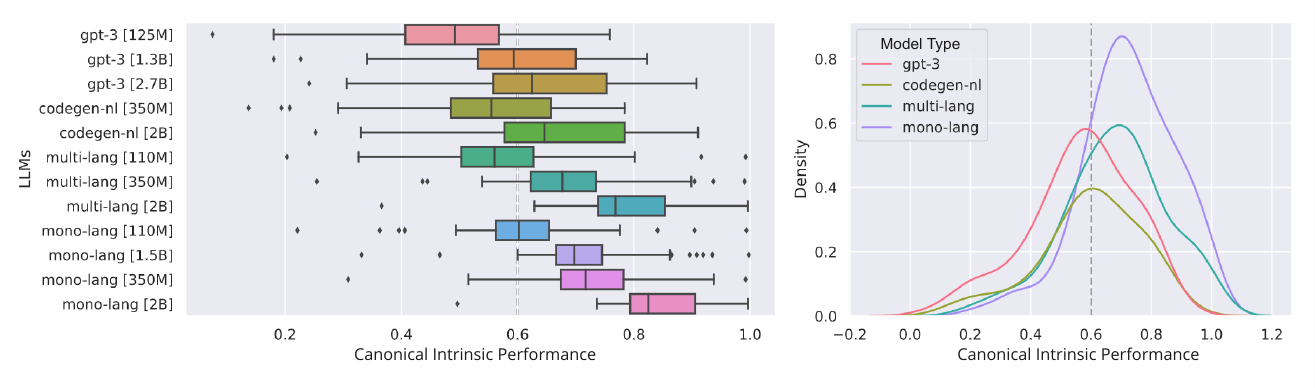}
\vspace{-1em}
\caption{{Canonical intrinsic performance for the models $M1$ to $M12$. Left: box-plots of performance distribution for each model. Right: density plot of performance by model type. }}
\Description{Canonical performance for each model under evaluation}
\label{fig:canonical_performance}        
\end{figure}

\begin{figure}[h]
		\centering
  \vspace{-1em}
		\includegraphics[width = 0.95\textwidth]{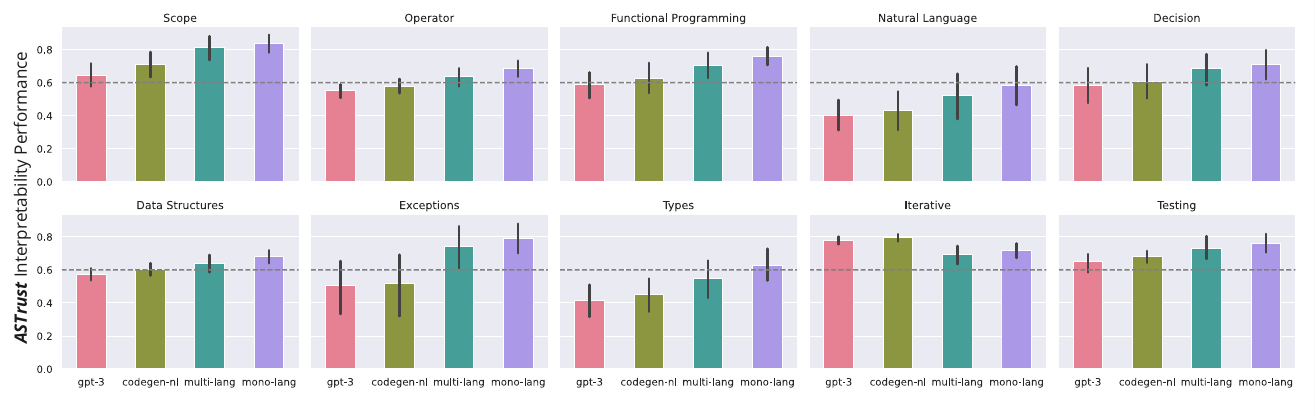}
  \centering
		\caption{Segregated \approach confidence by Syntax Categories (dotted line is the performance threshold).}
        \label{fig:asc_performance}     
        \vspace{-1em}
\end{figure}

{To assess the prediction confidence of each Syntax Category (\asc) for the 12 \llms we present an empirical \approach interpretability performance value (bootstrapped median columns in \tabref{tab:models_performance}). \figref{fig:asc_performance} illustrates the \approach interpretability performance segregated by Syntax Categories (\ascs) for each model type. Similarly, \tabref{tab:models_performance} shows bootstrapped median for each model. We set a confidence prediction threshold of $>=0.6$ across all analyses. It is worth noting that this threshold is a tunable parameter that can be modified to obtain tailored interpretations of model performance. We easily identify that \monolang and \multilang surpass our confidence threshold of $0.6$ on all the \ascs but \textit{Data Types}. Conversely, we observe that \textit{GPT-3-type} models face challenges in \textit{Data Types} categories while excelling in \textit{Iteration} categories.}
\begin{table}[h]
\centering
\caption{Syntax Concept Empirical Evaluation Results (bold: best, underlined: worst). }
\vspace{-0.2cm}
\label{tab:models_performance}

\scalebox{0.65}{%
\setlength{\tabcolsep}{5pt}

\begin{tabular}{clcccccccccclccc}
\hline
\multicolumn{1}{l}{} &  & \multicolumn{10}{c}{\textbf{\approach Interpretable Performance (bootstrapped median)}} &  & \multicolumn{2}{c}{\textbf{Extrinsic}} & \multicolumn{1}{l}{\textbf{Intrinsic}} \\ \cline{3-12} \cline{14-16} 
\multicolumn{1}{l}{\multirow{-2}{*}{\textbf{LLMs}}} &  & \multicolumn{1}{l}{\textit{Data Str.}} & \multicolumn{1}{l}{\textit{Decision}} & \multicolumn{1}{l}{\textit{Except.}} & \multicolumn{1}{l}{\textit{F. Prog.}} & \multicolumn{1}{l}{\textit{Iter.}} & \multicolumn{1}{l}{\textit{NL}} & \multicolumn{1}{l}{\textit{Oper.}} & \multicolumn{1}{l}{\textit{Scope}} & \multicolumn{1}{l}{\textit{Testing}} & \multicolumn{1}{l}{\textit{Data Ts.}} &  & \multicolumn{1}{l}{\textit{BLEU-4}} & \multicolumn{1}{l}{\textit{CodeBLEU}} & 
\textit{Perf.} \\ \hline
$M_1$ &  & 0.50 & 0.52 & \cellcolor[HTML]{FFCCC9}{\color[HTML]{000000} {\ul 0.43}} & \cellcolor[HTML]{FFCCC9}{\color[HTML]{000000} {\ul 0.49}} & 0.74 & \cellcolor[HTML]{FFCCC9}{\color[HTML]{000000} {\ul 0.32}} & \cellcolor[HTML]{FFCCC9}{\color[HTML]{000000} {\ul 0.48}} & 0.51 & 0.59 & \cellcolor[HTML]{FFCCC9}{\color[HTML]{000000} {\ul 0.33}} &  & 0.013 & 0.139 & 0.48 \\
$M_2$ &  & 0.60 & 0.61 & 0.53 & 0.62 & 0.79 & \cellcolor[HTML]{FFCCC9}{\color[HTML]{000000} {\ul 0.43}} & 0.57 & 0.68 & 0.68 & \cellcolor[HTML]{FFCCC9}{\color[HTML]{000000} {\ul 0.44}} &  & 0.016 & 0.151 & 0.59 \\
$M_3$ &  & 0.62 & 0.63 & 0.56 & 0.66 & \cellcolor[HTML]{CACDFA}\textbf{0.81} & \cellcolor[HTML]{FFCCC9}{\color[HTML]{000000} {\ul 0.46}} & 0.60 & 0.74 & 0.70 & \cellcolor[HTML]{FFCCC9}{\color[HTML]{000000} {\ul 0.47}} &  & 0.015 & 0.163 & \textbf{0.62} \\
$M_4$ &  & 0.56 & 0.57 & \cellcolor[HTML]{FFCCC9}{\color[HTML]{000000} {\ul 0.45}} & 0.57 & 0.77 & \cellcolor[HTML]{FFCCC9}{\color[HTML]{000000} {\ul 0.39}} & 0.54 & 0.64 & 0.64 & \cellcolor[HTML]{FFCCC9}{\color[HTML]{000000} {\ul 0.40}} &  & 0.015 & 0.151 & 0.55 \\
$M_5$ &  & 0.65 & 0.65 & 0.58 & 0.68 & \cellcolor[HTML]{CACDFA}\textbf{0.82} & \cellcolor[HTML]{FFCCC9}{\color[HTML]{000000} {\ul 0.48}} & 0.61 & 0.78 & 0.72 & 0.50 &  & 0.016 & 0.155 & \textbf{0.65} \\
$M_6$ &  & 0.54 & 0.55 & 0.64 &  0.60 & {\ul 0.60} & \cellcolor[HTML]{FFCCC9}{\color[HTML]{000000} {\ul 0.40}} & 0.54 & 0.71 & 0.67 & \cellcolor[HTML]{FFCCC9}{\color[HTML]{000000} {\ul 0.42}} &  & 0.010 & 0.189 & 0.57 \\
$M_7$ &  & 0.63 & 0.72 & 0.75 & 0.70 & 0.69 & 0.51 & 0.62 & 0.83 & 0.73 & 0.51 &  & 0.015 & 0.171 & 0.68 \\
$M_8$ &  & 0.74 & 0.79 & \cellcolor[HTML]{CACDFA}\textbf{0.83} & 0.81 & 0.77 & 0.65 & 0.74 & \cellcolor[HTML]{CACDFA}\textbf{0.91} & 0.80 & 0.71 &  & 0.016 & 0.177 & \textbf{0.79} \\
$M_9$ &  & 0.58 & 0.58 & 0.68 & 0.66 & 0.63 & \cellcolor[HTML]{FFCCC9}{\color[HTML]{000000} {\ul 0.46}} & 0.57 & 0.73 & 0.69 & \cellcolor[HTML]{FFCCC9}{\color[HTML]{000000} {\ul 0.47}} &  & 0.011 & 0.194 & 0.61 \\
$M_{10}$ &  & 0.67 & 0.67 & \cellcolor[HTML]{CACDFA}\textbf{0.80} & 0.76 & 0.70 & 0.59 & 0.66 & \cellcolor[HTML]{CACDFA}\textbf{0.82} & 0.74 & 0.64 &  & 0.011 & 0.196 & 0.71 \\
$M_{11}$ &  & 0.68 & 0.76 & 0.78 & 0.76 & 0.73 & 0.57 & 0.68 & \cellcolor[HTML]{CACDFA}\textbf{0.86} & 0.77 & 0.58 &  & 0.014 & 0.179 & 0.73 \\
$M_{12}$ &  & 0.79 & \cellcolor[HTML]{CACDFA}\textbf{0.84} & \cellcolor[HTML]{CACDFA}\textbf{0.90} & \cellcolor[HTML]{CACDFA}\textbf{0.85} & \cellcolor[HTML]{CACDFA}\textbf{0.81} & 0.73 & \cellcolor[HTML]{CACDFA}\textbf{0.82} & \cellcolor[HTML]{CACDFA}\textbf{0.94} & \cellcolor[HTML]{CACDFA}{\color[HTML]{000000} \textbf{0.85}} & \cellcolor[HTML]{CACDFA}\textbf{0.83} &  & 0.016 & 0.181 & \textbf{0.84} \\ \hline
\end{tabular}
}
\vspace{0.1cm}
{\\ \footnotesize *Erroneous \approach values are in red. Confident \approach scores are in blue. Canonical values serve as a baseline.}
\vspace{-0.3cm}

\end{table}

{We found that categories such as \textit{Iteration}, \textit{Except}, and \textit{Scope} surpass our threshold for the majority of our \llms under analysis. For instance, \tabref{tab:models_performance} shows the \textit{Iteration} category consistently surpasses our threshold for all \llms, except for \multiI ($M_6$), which records an average median \approach of $0.6$ and a highest value of $0.82$ for \codegenII ($M_5$). Notably, our smaller model \gptI ($M_1$) still outperforms the \textit{Iteration} category prediction with an average median of $0.74$. Finally, we note that models trained largely on code, i.e. \codegenII ($M_5$), \monoII ($M_{10}$), \monoIIII ($M_{12}$), could predict the \textit{Data Types} category with and \approach performance of $0.71$, $0.64$ and $0.83$ respectively.}

By contrast, our \llms struggle to generate good predictions for \textit{Natural language} and \textit{Data Types} categories. We can observe that only \monoIIII ($M_{12}$) surpasses our threshold for \textit{Natural language} with confidence of $0.73$, which is still not an outstanding probability. We attribute poor \approach performance in certain models to the nature of syntax categories like \textit{Natural Language} and \textit{Data Types}, which demand a larger input context for accurate prediction

\begin{figure}[h]
\centering
\includegraphics[width=0.9\textwidth]{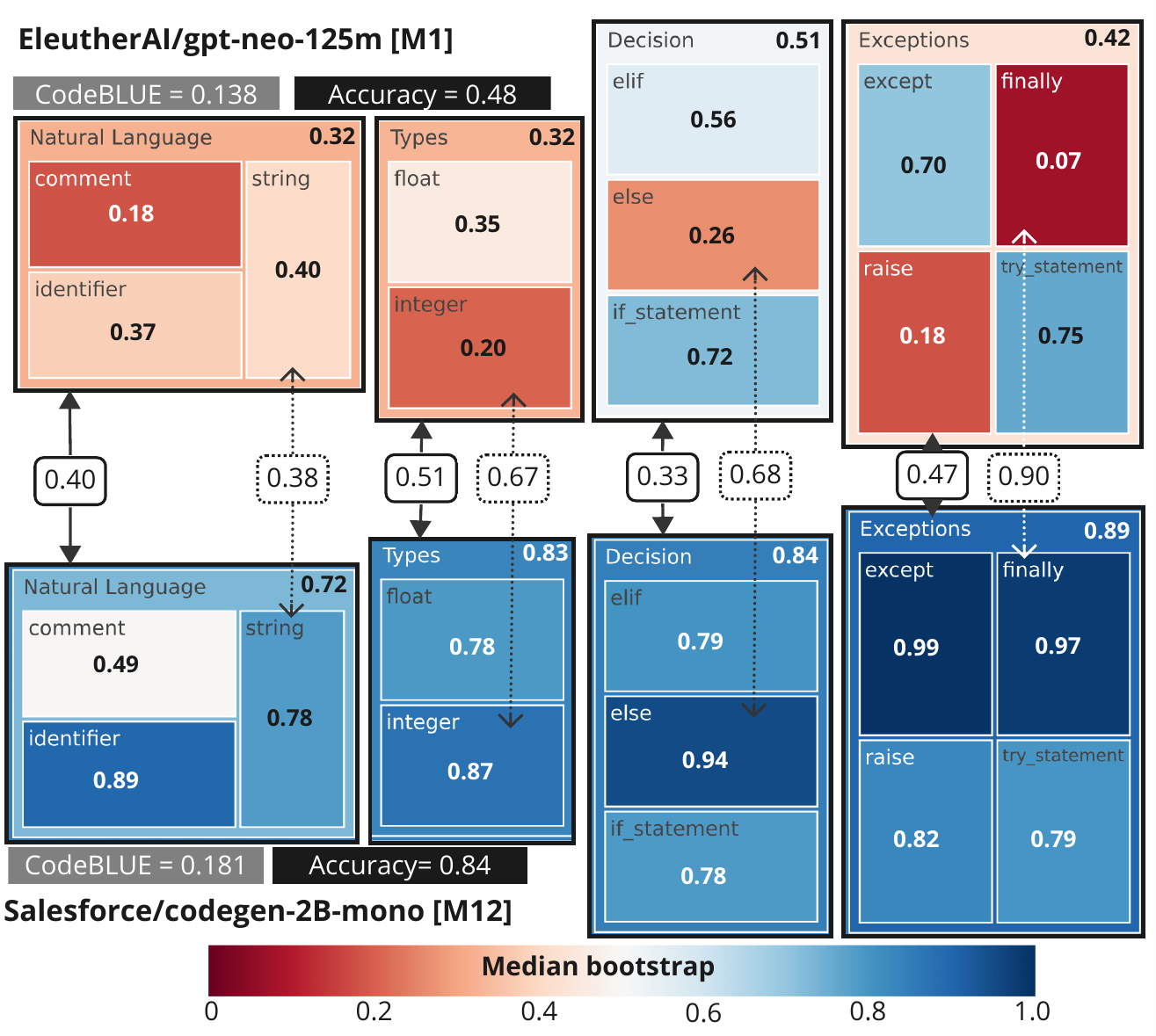}
\caption{\approach of Syntax Categories and Subcategories (dotted boxes) for corner cases ($M_1$ and $M_{12}$)}
\vspace{-1em}
\label{fig:largeTreeMap}
\Description{Syntax Categories heatmap comparing smaller and largest models}
\end{figure}

Our observations indicate that scaling \llms' parameters positively influences the prediction of \ascs. This scaling observation is consistent with canonical scores since our largest models \gptIII ($M_3$), \codegenII ($M_5$), \multiIII ($M_{8}$), and \monoIIII ($M_{12}$) report not only intrinsic accuracy that surpasses our threshold but also \approach confidence over $0.8$ for categories such as \textit{Exception}, \textit{Iteration}, and \textit{Scope} (see \tabref{tab:models_performance} in blue).  For instance, the largest model, $M_{12}$ exhibits the highest intrinsic accuracy with an avg. median of $0.84$ and exceeds our threshold for each category.

By comparing our \approach against extrinsic metrics, we observe that $M_{12}$ does not achieve the highest \textit{CodeBLUE} score, recording a $0.181$. Thus, \approach offers additional insights into the performance of syntax categories. For example, while \monoI ($M_{1}$) outperforms \monoIIII ($M_{12}$) with a \textit{CodeBLUE} of $0.194$, it struggles with inferring \textit{Natural Language} and \textit{Data Types} categories with $0.46$ and $0.47$ respectively (see \tabref{tab:models_performance}). 

{\textit{Corner Case Experiment.} \figref{fig:largeTreeMap}, shows a heatmap with the smallest and largest \llms under analysis, \gptI ($M_1$) and \monoIIII ($M_{12}$) respectively. We selected subcategories and categories with the greatest score jumps for going further into the \asc analysis. For instance, \textit{Data Types} reported the biggest difference with $0.51$ meanwhile the difference between subcategories such as  \texttt{\small `string'} and \texttt{\small `finally'} are $0.38$ and $0.9$, respectively. This difference suggests that model scaling positively impacts the \texttt{\small `finally'} subcategory, while \texttt{\small `string'} or \texttt{\small `if\_statement'} subcategories are slightly affected by the model size. We hypothesize that the poor performance of the \textit{Natural Language} is due to limited context windows in the prompt to predict this category. However, a complementary large-scale exploratory analysis of the proportionality types in the training and testing data is required beforehand to determine other causes of poor performance. We also observe that the \textit{Data Types} is prone to errors, especially as these types may frequently appear at the beginning of code snippets, particularly in Python, where dynamic typing is prevalent. This inter-comparison (\aka across models) would have been infeasible by just using canonical accuracy metrics. }

\begin{boxK}
\textit{\ref{rq:performance}:}
\approach allows us to segregate the prediction performance of \llms according to Syntax Categories, showing a more interpretable way of comparing models. Syntax-grounded explanations demonstrate, for instance, the struggle of \llms to statistically learn Natural Language nested within code structures. 
\end{boxK}

\subsection{\ref{rq:validity} \approach Validity}~\label{sec:global_validity}
\vspace{-1em}

We quantitatively demonstrate that cross-entropy loss of \llms tends to be negatively impacted by \approach probabilities. Therefore, we can explain at syntax category granularity which parts of the code \llms perform poorly (see red boxes in \tabref{tab:models_performance}). We showcase empirical evidence that the previous statement holds for correlations $\rho$ and causal effects $p(y|do(t))$. \tabref{tab:correlations} shows, in general, \ascs (\eg Iterative, Scope, or Operator) negatively influence the cross-entropy loss for our best (\ie $M_{12}$) and worst (\ie $M_{1}$) models. Negative effects indicate that the better a syntax category is predicted, the lower the learning error associated.

\begin{table}[ht]
\centering
\caption{\approach influence on Learning Error.}
\vspace{-0.2cm}
\label{tab:correlations}

\scalebox{0.85}{

\setlength{\tabcolsep}{4pt} 

\begin{tabular}{lllrrrr}
\hline
\multicolumn{2}{c}{\textbf{{[}T{]} Syntax Categories}} &  & \multicolumn{4}{c}{\textbf{{[}Y{]} Learning Error}} \\ \cline{1-2} \cline{4-7}
\multicolumn{1}{c}{\textbf{Categories}} & \multicolumn{1}{c}{\textbf{Sub-Categories}} &  & \multicolumn{2}{c}{\textbf{gpt-125}} & \multicolumn{2}{c}{\textbf{mono-2B}} \\  \cline{4-7}
\multicolumn{1}{c}{\textbf{$\mathcal{C}$}} & \multicolumn{1}{c}{\textbf{$\alpha,\lambda$}} &  & \multicolumn{1}{c}{$\rho$} & \multicolumn{1}{c}{\textbf{ATE}} & \multicolumn{1}{c}{$\rho$} & \multicolumn{1}{c}{\textbf{ATE}} \\ \hline
 & \textit{for\_statement} &  & -0.16 & \cellcolor[HTML]{EFEFEF}-0.10 & -0.07 & \cellcolor[HTML]{EFEFEF}-0.01 \\
\multirow{-2}{*}{\textit{\textbf{Iterative}}} & \textit{while\_statement} &  & -0.05 & \cellcolor[HTML]{EFEFEF}-0.11 & -0.03 & \cellcolor[HTML]{EFEFEF}-0.08 \\ \hline
 & \textit{identifier} &  & \textbf{-0.56} & \cellcolor[HTML]{EFEFEF}\textbf{-1.78} & \textbf{-0.80} & \cellcolor[HTML]{EFEFEF}\textbf{-2.89} \\
\multirow{-2}{*}{\textit{\textbf{\begin{tabular}[c]{@{}l@{}}Natural \\ Language\end{tabular}}}} & \textit{string} &  & -0.31 & \cellcolor[HTML]{EFEFEF}-0.36 & \textbf{-0.43} & \cellcolor[HTML]{EFEFEF}\textbf{-0.55} \\ \hline
 & \textit{return\_statement} &  & -0.04 & \cellcolor[HTML]{EFEFEF}-0.09 & -0.22 & \cellcolor[HTML]{EFEFEF}-0.09 \\
 & \textit{{]}} &  & -0.16 & \cellcolor[HTML]{EFEFEF}-0.04 & -0.22 & \cellcolor[HTML]{EFEFEF}-0.10 \\
\multirow{-3}{*}{\textit{\textbf{Scope}}} & \textit{)} &  & -0.37 & \cellcolor[HTML]{EFEFEF}-0.85 & \textbf{-0.54} & \cellcolor[HTML]{EFEFEF}\textbf{-1.49} \\ \hline
\textit{\textbf{Decision}} & \textit{if\_statement} &  & -0.22 & \cellcolor[HTML]{EFEFEF}-0.21 & -0.11 & \cellcolor[HTML]{EFEFEF}-0.11 \\ \hline
 & \textit{comparison\_operator} &  & -0.13 & \cellcolor[HTML]{EFEFEF}0.02 & -0.11 & \cellcolor[HTML]{EFEFEF}0.00 \\
\multirow{-2}{*}{\textit{\textbf{Operator}}} & \textit{boolean\_operator} &  & -0.10 & \cellcolor[HTML]{EFEFEF}0.01 & -0.08 & \cellcolor[HTML]{EFEFEF}-0.09 \\ \hline
 & \textit{for\_in\_clause} &  &  -0.03 & \cellcolor[HTML]{EFEFEF}0.09 & -0.03 & \cellcolor[HTML]{EFEFEF}0.04 \\
 & \textit{if\_clause} &  & -0.01 & \cellcolor[HTML]{EFEFEF}0.19 & 0.01 & \cellcolor[HTML]{EFEFEF}0.13 \\
 & \textit{lambda} &  & -0.04 & \cellcolor[HTML]{EFEFEF}0.20 & -0.05 & \cellcolor[HTML]{EFEFEF}0.06 \\
\multirow{-4}{*}{\textit{\textbf{\begin{tabular}[c]{@{}l@{}}Functional \\ Programming\end{tabular}}}} & \textit{list\_comprehension} &  & -0.04 & \cellcolor[HTML]{EFEFEF}0.06 & -0.03 & \cellcolor[HTML]{EFEFEF}0.04 \\ \hline
\multicolumn{2}{c}{\textbf{Baseline}} &  & \multicolumn{1}{l}{} & \multicolumn{1}{l}{} & \multicolumn{1}{l}{} & \multicolumn{1}{l}{} \\ \cline{1-2}
\textbf{Intrinsic} & Avg. Accuracy &  & -0.38 & \cellcolor[HTML]{EFEFEF}\textbf{-1.60} & -0.38 & \cellcolor[HTML]{EFEFEF}\textbf{-1.78} \\ \hline
\end{tabular}
} %
\vspace{0.1cm}
{\\ \footnotesize{* \textbf{bold}:Highest impact, shadowed: causal effect}}
\end{table}

The most outstanding finding is that the \textit{Natural Language} category has the largest impact on the cross-entropy loss. For example, the \asc \texttt{\small `identifier'} has a causal effect of $-1.78$ for $M_{1}$ and $-2.89$ for $M_{12}$. In contrast, \textit{Functional Programming} categories present the lowest impact on cross-entropy loss with a subtle \texttt{\small `lambda'} positive causal effect of $0.2$ for $M_{1}$. This subtle positive effect was expected as NL-based \llms have not been fine-tuned on code corpora with \texttt{\small `lambda'} expressions. 

\begin{boxK}
\textit{\ref{rq:validity}: }
The cross-entropy loss of \llms tends to be negatively impacted by \approach probabilities. This demonstrates that syntax-grounded explanations are indeed representing the syntax learning mechanisms of \llms at segregated granularity. 
\end{boxK}

\section{Discussion}\label{sec:discussion}

{Below, we pose three aspects for discussion: 1) some general insights ($GIs$) from the empirical study, 2) a logical analysis of the connection between trustworthiness and interpretability, and 3) the threats to validity of our approach.}

\subsection{Empirical Study Insights}
\textbf{$GI_1$: Token Predictions Reliability.} { \approach relies on logit extraction to generate post-hoc explanations as syntax categories. If logits are wrongly predicted (by over/underfitting), our \textit{causal validation process} detects such inconsistency by reducing the Average Treatment Effect (ATE) of syntax categories on the statistical learning error. Our Structural Causal Model (SCM) was designed to test the robustness and fidelity of our approach under models’ misconfigurations or unreliable performance. Also, as stated earlier in the paper, recent work on calibration for LLMs of code has illustrated that, for code completion (which subsumes the experimental settings in this paper), LLMs tend to be well calibrated to token probabilities/logits~\cite{spiess2024quality}. This helps to mitigate issues that may arise due to model confidence and correctness being misaligned.}

\textbf{$GI_2$: Syntax Aggregations Improves Explanations.} {Due to its granular nature, token-level predictions are less informative than a hierarchical aggregated level. BPE can make the interpretation of individual tokens much more difficult when code-based sequences are split into tokens that may be meaningless. We posit that practitioners can more easily understand syntax categories rather than individual tokens because these categories are \textit{already defined by context-free grammars}, which are semantically rich. Moreover, our human study provides evidence of this claim since AST-based explanations were found to be easy to read and use by participants. AST-based explanations \textit{also capture semantics} by allowing visualization of the full AST structure. This approach helps practitioners evaluate the model's implementation more effectively by providing a clearer, structured view of the code's semantics.}

\textbf{$GI_3$: Natural Language Imbalance.} {Our approach indicates a poor performance on NL sub-categories. We hypothesize this low performance is due to an unbalanced distribution of NL training samples compared to other categories. Before increasing the context window, we believe that a better analysis would be measuring the proportionality of NL sub-categories on the training set and, then, fine-tuning current LLMs to fix possible data bias. Unfortunately, this analysis is currently out of scope since it demands a complementary Exploratory Data Analysis that we envision for future research stages.}

\textbf{$GI_4$: Foundational Interpretability Research.} {\approach is meant to serve as a more foundational approach required to guide the future development of interpretability tools for users of different backgrounds (\eg researchers, students, and data scientists). We aimed to not only propose a formal methodology to conduct interpretability in our field but also perform a preliminary assessment of \approach’s usefulness by conducting a control/treatment experiment (\ie with and without the approach) on a visualization technique based on \approach under clearly defined qualitative metrics.}

\textbf{$GI_5$: Contradictions about the Usefulness of Explanations.} {In our human study, we found that AST-based explanations were preferred over sequential-based ones. Results revealed that the AST-partial representation was considered more useful than AST-Complete, as it presents the AST representation and \approach confidence performance only for the generated portion of the code. However, the feedback received in the open-ended questions revealed contradictory opinions. Some participants indicated that the AST-partial representation missed important details, while others felt that the AST-Complete representation was excessively detailed. These findings suggest the need for more tailored representations for explanations, aiming to present useful information while maintaining readability. We envision incorporating \approach into a tool that adds interactivity to navigate the explanations.}

\label{sec:review}

\subsection{Trustworthiness \& Interpretability Connection}
\label{sec:logical}
We outline two premises based on state-of-the-art definitions of trustworthiness and direct observations from our quantitative and qualitative analyses. Then, we use \textit{logical deduction} supported by documented and empirical evidence to link the concept of \textit{trustworthiness} with \textit{\approach} highlighting the significance of syntax-grounded explanations.

\textbf{Premise$_1$: Interpretability is a cornerstone for trustworthiness in Language Models for Code (LLMs)}. The interpretability field enhances transparency and provides insights into the decision-making process serving as a key factor in fostering practitioner trust and adoption. In the realm of Deep Learning for Software Engineering (DL4SE) ~\cite{watson2021systematic}, the significance of interpretability in models to engender trust cannot be overstated. Jiaming \etal ~\cite{ji2024ai} underscore interpretability as a pivotal element in aligned models, integral to the RICE principle alongside Robustness, Controllability, and Ethicality. By enhancing transparency in the decision-making process, interpretability plays a crucial role in building trust, a sentiment echoed by Weller \etal, who stress the need to extend transparency beyond algorithms to foster trust ~\cite{weller2019transparency}. The dilemma between accuracy and interpretability, claimed by Lundberg \etal ~\cite{lundberg2017unified}, is magnified by the challenge posed by large, complex models that even experts find difficult to interpret. A user study with UX and design practitioners supports this notion, revealing that explanations are sought to gain deeper insights into AI tool decision-making, providing a remedy for the ``black box'' perception and contributing to user trust and adoption ~\cite{Liao_2020}. Therefore, the indisputable importance of interpretability in \llm decision-making lies in its pivotal role in establishing trustworthiness.

\textbf{Premise$_2$: \approach improves interpretability.} It is feasible to segregate intrinsic metrics (\ie standard accuracy) into interpretable Syntax Categories revealing the \llms' inner workings concerning code structure and contributing towards interpretability. By conducting extensive qualitative and quantitative studies involving 12 prominent \llms, we have demonstrated the effectiveness of \approach in enhancing interpretability.  We do not claim that our set of categories is complete; however, we consider that a good \textit{alignment} of the generated categories by the \llm with the ones expected by humans configures a good explanation~\cite{ghorbani_towards_2019}. Our \approach clusters tokens to \textit{meaningful} categories that are easier for human concept association. Furthermore, we uncovered valuable insights, such as the causal influence of AST categories on the cross-entropy loss of \llms after accounting for confounding factors. Our human study participants attested to the usefulness of our \approach in explaining the predictions of Python code snippets by a \llm \ref{sec:local1}. By breaking down intrinsic metrics into segregated and interpretable terminal and non-terminal nodes, our approach not only enhances the understandability of \llms but also unveils crucial insights into the inner workings of syntax elements.

\textbf{Conclusion.} Given the first premise that interpretability is fundamental for trustworthiness in \llms, supported by several shreds of evidence, and from the second premise asserting that \approach enhances interpretability by segregating intrinsic metrics into interpretable syntax categories and subcategories, collectively supports the fact that \approach contributes to the improvement of trustworthiness in \llms for Code using syntax-grounded explanations. 

\subsection{Threats to Validity} \label{sec:threats}

Threats to \textbf{construct validity} concern the intentionality of \approach in providing useful explanations. Instead of attempting to disentangle information represented between the layers learned by \llms (\ie probing ~\cite{lopez_ast-probe_2022}), \approach focuses on conceptually mapping \llms' code predictions to present the accuracy in a segregated way. We quantitatively and qualitatively validated the extent to which \approach is interpretable through causal analyses and a human study.
{While we cannot claim that the results from our study generalize beyond the population of users that participated in our study, our participants represent a diverse range of backgrounds mitigating this threat.  Nonetheless, the purpose of our study was to conduct a preliminary assessment of \approach representations. As such, all code completion scenarios were designed to include a syntax or semantic error since we assessed how useful our approach is in assisting users in understanding models' incorrect behavior, increasing the reliability of our findings.}

Threats to \textbf{internal validity} refer to the degree of confidence in which the \approach study results are reliable. Firstly, in our causal study, the potential for unidentified confounders in the code may bias the causal relationship between cross-entropy loss and the Syntax Categories. That is why we ensured the robustness of the Structural Causal Model by performing placebo refutations, which involves simulating unrelated treatments and then re-estimating the causal effects. Secondly, we used rigorous statistical techniques such as bootstrapping to guarantee a consistent comparison between aggregated Token-Level Predictions by syntax elements. 

Threats to \textbf{external validity} represent the extent to which \approach can be used to contextualize the performance of other \llms or datasets. We excluded GPT-4 based models from our empirical experiments due to the constraints of the current OpenAI API, which restricts access to softmax layers values (a key factor in the intrinsic evaluations). %
While our evaluation relied on decoder-only based models, \approach can also be used to interpret encoders and other types of auto-regressive architectures. Finally, our \galeras dataset may not contain enough samples to represent all the syntax categories or Python project attributes fairly. Nonetheless, we designed a data mining pipeline to guarantee the diversity of collected samples.

\section{Lessons Learned \& Conclusions}\label{sec:conclusion}

\textbf{Lesson$_1$: Aggregated metrics may give false impressions about \llms' capabilities.} The research community should incentivize researchers to report AI4SE results in a granular way, as opposed to more traditional aggregated accuracy metrics. After controlling for code confounders, we demonstrated that segregated syntax elements influence the cross-entropy loss of \llms. This influence persists across models at different parameter sizes and fine-tuning strategies. Syntax information is also relevant for any posterior static analysis of code enabling further evaluations of \llms in downstream tasks that entail elements of software design (\eg refactoring).

\textbf{Lesson$_2$: New interpretability methods are required to enable trustworthiness.} In our studies, we have noted an absence of a concrete definition for the term \textit{trust} in the Software Engineering research. However, several researchers have highlighted the importance of establishing trust in work on AI4SE. Research has also shown that interpretability is one of the keys to improving trustworthiness, but at the same time, there is a scarcity of interpretable methods linked to trustworthiness. Despite this limitation, surveyed participants agreed that \approach was useful to understand why and how a \llm produced certain errors in code-completion tasks. 

\textbf{Lesson$_3$: Grounding model explanations in the relationship between syntactic structures and prediction confidence is useful.} It is feasible to segregate intrinsic metrics (\ie standard accuracy) into interpretable Syntax Categories revealing the \llms' inner workings concerning code structure and contributing towards interpretability. By conducting extensive qualitative and quantitative studies involving 12 prominent \llms, we have demonstrated the effectiveness of \approach in enhancing interpretability.  We do not claim that our set of categories is complete; however, we consider that a good \textit{alignment} of the generated categories by the \llm with the ones expected by humans configures a good explanation~\cite{ghorbani_towards_2019}. Our \approach clusters tokens to \textit{meaningful} categories that are easier for human concept association. Furthermore, we uncovered valuable insights, such as the causal influence of AST categories on the cross-entropy loss of \llms after accounting for confounding factors. Our human study participants attested to the usefulness of our \approach in explaining the predictions of Python code snippets by a \llm \ref{sec:local1}. By breaking down intrinsic metrics into segregated and interpretable terminal and non-terminal nodes, our approach not only enhances the understandability of \llms but also unveils crucial insights into the inner workings of syntax elements.

\textbf{Lesson$_4$: The usability of proposed techniques must be further evaluated for industry adoption.} We adapted the non-mathematical definition of \textit{interpretability} by Doshi-Velez \& Kim ~\cite{doshi-velez_towards_2017}, Molnar ~\cite{molnar_interpretable_2020} and Miller ~\cite{miller2018explanation} to the field of AI4SE~\cite{watson2021systematic}. However, as our preliminary human study suggests, \approach solution is incomplete until being extensively evaluated for industry settings. 

\noindent\textbf{Artifact Availability:} Experimental data, curated datasets, source code, and complementary statistical analysis used in this research are published in an open-source repository~\cite{AnonyRepoASTrust24}.

\bibliographystyle{ACM-Reference-Format-num}
\bibliography{utils/rationales,utils/ccp}

\end{document}